# Performance Analysis of Robust Stable PID Controllers Using Dominant Pole Placement for SOPTD Process Models

Saptarshi Das[1], Kaushik Halder[2], and Amitava Gupta[2]

1) Department of Mathematics, College of Engineering, Mathematics and Physical Sciences, University of Exeter, Penryn Campus, Penryn TR10 9FE, United Kingdom
2) Department of Power Engineering, Jadavpur University, Salt Lake Campus, LB-8, Sector 3, Kolkata-700098, India

**Author's Emails:**

s.das3@exeter.ac.uk, saptarshi@pe.jusl.ac.in (S. Das*)

khalder.pe@research.jusl.ac.in (K. Halder)

amitg@pe.jusl.ac.in (A. Gupta)

**Phone number**: +44-7448572598

**Abstract:**

This paper derives new formulations for designing dominant pole placement based proportional-integral-derivative (PID) controllers to handle second order processes with time delays (SOPTD). Previously, similar attempts have been made for pole placement in delay-free systems. The presence of the time delay term manifests itself as a higher order system with variable number of interlaced poles and zeros upon Pade approximation, which makes it difficult to achieve precise pole placement control. We here report the analytical expressions to constrain the closed loop dominant and non-dominant poles at the desired locations in the complex *s*-plane, using a third order Pade approximation for the delay term. However, invariance of the closed loop performance with different time delay approximation has also been verified using increasing order of Pade, representing a closed to reality higher order delay dynamics. The choice of the nature of non-dominant poles e.g. all being complex, real or a combination of them modifies the characteristic equation and influences the achievable stability regions. The effect of different types of non-dominant poles and the corresponding stability regions are obtained for nine test-bench processes indicating different levels of open-loop damping and lag to delay ratio. Next, we investigate which expression yields a wider stability region in the design parameter space by using Monte Carlo simulations while uniformly sampling a chosen design parameter space. The accepted data-points from the stabilizing region in the design parameter space can then be mapped on to the PID controller parameter space, relating these two sets of parameters. The widest stability region is then used to find out the most robust solution which are investigated using an unsupervised data clustering algorithm yielding the optimal centroid location of the arbitrary shaped stability regions. Various time and frequency domain control performance parameters are investigated next, as well as their deviations with uncertain process parameters, using thousands of Monte Carlo simulations, around the robust stable solution for each of the nine test-bench processes. We also report, PID controller tuning rules for the robust stable solutions using the test-bench processes while also providing computational complexity analysis of the algorithm and carry out hypothesis testing for the distribution of sampled data-points for different classes of process dynamics and non-dominant pole types.

**Nomenclature:**

$m$ : Non-dominance parameter for pole placement,

$K$ : Open loop process DC gain,

$L$ : Open loop process time delay,

$T$ : Open loop process time constant or lag,

$\omega_{ol} = 1/T$ : Open loop process natural frequency,

$\xi_{ol}$ : Open loop process damping ratio,

$\omega_{cl}$ : Closed loop process natural frequency,

$\zeta_{cl}$ : Closed loop process damping ratio,

$K_p$ : Proportional gain,

$K_i$ : Integral gain,

$K_d$ : Derivative gain,

$s$ : Laplace operator,

$\mathcal{H}_2/\mathcal{H}_\infty$ : System norms,

$\mathcal{L}_2/\mathcal{L}_\infty$ : Signal norms,

$G_m$ : Gain margin,

$\Phi_m$ : Phase margin,

$\omega_{gc}$ : Gain crossover frequency,

$d(t)$: Disturbance input,

$u(t)$: Control signal,

$y(t)$: output variable,

$e(t)$: control error signal,

$J$: Performance measure,

$S^e(s)$ : sensitivity function,

$T(s)$ : complementary sensitivity function,

$S^d(s)$ : disturbance sensitivity function,

$S^u(s)$ : control sensitivity function

1. **Introduction**

The dynamic behaviours of many industrial processes are affected and governed by significant amount of time delays in the control loops. The time delay is caused by the flow of information, energy and transport of physical variables, computer processing time etc. (Normey-Rico & Camacho 2007). The introduction of the time delay makes the continuous time closed loop system to have an infinite order (Åström & Wittenmark 2011) upon exponential series expansion of the delay term ( $e^{-Ls}$ ) which is difficult to handle with a finite term controller (Michiels et al. 2002). To alleviate this problem, there have been several works to design controllers for such systems e.g. in (Zhong 2006). It is well known that most of the controllers used in the process industries are of PID type (Åström & Hägglund 1995) (Zhong 2006) due to its simplicity and ease of implementation, nice disturbance rejection, tracking performance etc. Amongst many other available approaches, the Internal Model Control (IMC) based tuning of PID controllers has been quite popular to handle First Order Plus Time Delay (FOPTD) and Second Order Plus Time Delay (SOPTD) processes, as well as Integral Process with Dead Time (IPDT) (Shamsuzzoha & Lee 2007; Rivera et al. 1986) because of its good robustness on uncertain plants. Another approach on the design of smith predictor augmented PID controller to handle time delay processes have been reported in (Astrom et al. 1994) which yields improved tracking and load disturbance rejection performances. A modified methodology is proposed with combined Smith predictor and PID controller in (Matausek & Micic 1996) considering challenging higher order integrating plants with delays. However the main drawback of this method is that it cannot handle unstable process with delay (Normey-Rico & Camacho 2007), unless an additional observer is used (Furukawa & Shimemura 1983). To overcome these problems of complicated time delay processes, the model predictive control (MPC) has got attention by many researchers but there are only few results for time delay systems (Ellis & Christofides 2015). Initially the MPC was developed mainly to control slower processes as it requires large computational burden for prediction and optimisation-based control. MPC controller design for time delay systems is mostly an open area and there are only few results like (Holis & Bobál 2015). Another important area is designing output feedback controller (De la Sen 2005) as well as state feedback controller (Michiels et al. 2010) for stabilizing time delayed systems which are gradually gaining increased attention. In the literature some control algorithms are proposed using linear matrix inequalities (LMIs) for time delayed systems e.g. (Niculescu 1998) to enforce robustness and several closed loop performance measures. Despite having these results, traditional pole placement remained quite challenging for time delay systems because of its increased or even infinite order.

In this paper, we propose an analytical formulation for dominant pole placement tuning of PID controllers to handle SOPTD systems. This is due to the fact that in many process industries, the dynamical behaviours in a large variety of self-regulating processes can be modelled using the SOPTD template with the flexibility of showing both sluggish and oscillatory open loop dynamics as well as different lag to delay ratio or normalized dead time (O'Dwyer 2009). PID controllers are traditionally tuned by various means like time and frequency domain performance criteria or design specifications (Cominos & Munro 2002). Amongst many available approaches, the dominant pole placement method has been quite promising as the designer can specify his demand of closed loop performance, as the equivalent second order system's damping ratio, time constant or natural frequency (Wang et al. 2009). Amongst previous approaches, dominant pole placement based PID controller design for delay free second order systems have been addressed in (Saha et al. 2012)(Das, Halder, et al. 2012) whereas its time delay version has been extended in this paper and the method has been verified on several test-bench SOPTD plants.

A continuous pole placement method (controlling the rightmost root of the closed loop system and shift it to the left half of the *s*-plane in quasi-continuous way) has been proposed to design a semi-automated pole placement based state feedback controller for retarded (Michiels et al. 2002) and neutral type (Michiels & Vyhlidal 2005) delay systems. By this method, the closed loop roots lying in the extreme right-hand side is shifted to the far possible left-hand side. However, the methodology does not allow direct pole placement for SOPTD system and only monitoring the real part of the roots. To overcome the above problem another methodology is proposed in (Michiels et al. 2010) which combines direct pole placement and the minimization of the spectral abscissa for determining controller parameters in retarded time-delay systems. There are some other approaches on stability analysis of time delayed single input single output (SISO) systems to derive controller gains by computing the root locus. Using the characteristic equation which leads to a transcendental equation in the presence of delays which is also known as the quasi-polynomial, several methods have been proposed to construct the root locus which creates horizontal asymptotes (Krall 1970; Yeung & Wong 1982; Huang & Li 1967). Other root locus based stabilization methods are also reported to analyse state space models with input delay (Engelborghs et al. 2001), state delay or both (Suh & Bien 1982) by using the root locus.

The PID controller tuning by direct pole assignment is found to be a difficult approach for time delay systems, as the time delay in a process makes the closed loop system to have an infinite order. Therefore, to handle time delay systems, the direct pole placement in complex *s*-plane is not recommended as suggested in the pioneering work on dominant pole placement tuning in (Wang et al. 2009). The methodology in (Wang et al. 2009) also suggested a Nyquist based design for frequency domain stabilization of the time delay systems using PID controllers. The main hurdle with the pole placement for delay systems has been the fact that the exponential delay term in Laplace domain (i.e. $e^{-Ls}$), manifests itself as very high order transfer function upon approximations using Pade/Routh methods of a specified order (Silva et al. 2007). Therefore, such a pole placement approach will need relocation of many open loop poles at a time with a compact finite term (three-term for PID) controller when the number of interlacing pole-zeroes, arising due to the approximation of the time delay term are too many to handle for a chosen order of approximation. In our derivations, the order and approximation method are considered to be fixed. In particular, we apply a third order Pade approach to approximate the time delay term in the process model. Therefore, it can be considered as a new area of research to get a clearer picture of the dominant pole placement design for time delay systems where the task is to handle many poles and zeros of the combined plant and delay with finite number of controller parameters. To demonstrate the methodology, SOPTD processes with various delay to lag ratio have been used which shows the strength of the algorithm for even quite challenging plants e.g. delay dominant systems with low open loop damping which are much harder to control using standard PID controller tuning methods.

2. **Theoretical formulation**

In this section, the dominant pole placement based PID controller design has been shown to handle SOPTD systems. The time delay term has been approximated using a third order Pade's approximation instead of considering it as transcendental term in the quasi-polynomial (Silva et al. 2007). The co-efficient matching based pole placement method has been described previously in (Åström & Hägglund 1995)(Kiong et al. 1999) for the control of delay free systems. We apply here a similar co-efficient matching method to design dominant pole placement based PID controller gains using the characteristic equation having third order Pade approximation of the delay term.

Now, let us consider the open loop SOPTD process can be represented by:

$$G(s) = \frac{K}{s^2 + 2\zeta_{ol}\omega_{ol}s + \omega_{ol}^2} e^{-Ls},  \qquad (1)$$

which is to be controlled by the PID controller

$$C(s) = K_p + \frac{K_i}{s} + K_d s = \frac{K_d s^2 + K_p s + K_i}{s}. \qquad (2)$$

Then the closed loop system with PID controller can be written as:

$$G_{cl}(s) = \frac{G(s)C(s)}{1+G(s)C(s)}. \tag{3}$$

Again, using a third order Pade approximation for the time delay term ($e^{-Ls}$) of the open loop SOPTD process model (1) can be written as:

$$e^{-Ls} \approx \frac{-L^3s^3 + 12L^2s^2 - 60Ls + 120}{L^3s^3 + 12L^2s^2 + 60Ls + 120}. \tag{4}$$

Therefore, the full expression of the closed loop system transfer function including the open loop plant and PID controller is given by:

$$G_{cl} = \frac{K(K_d s^2 + K_p s + K_i)(-L^3s^3 + 12L^2s^2 - 60Ls + 120)}{s(s^2 + 2\zeta_{ol}\omega_{ol}s + \omega_{ol}^2)(L^3s^3 + 12L^2s^2 + 60Ls + 120) + K(K_d s^2 + K_p s + K_i)(-L^3s^3 + 12L^2s^2 - 60Ls + 120)}. \tag{5}$$

It is evident from (5) that the closed loop process has six poles and five zeros which can be modulated using suitable choice of the three controller gains $\{K_p, K_i, K_d\}$. As the closed loop system in (5) has six poles, the corresponding desired closed-loop characteristic polynomial should also contain six roots out of which two must meet the equivalent desired closed loop system specifications. Rest of the four non-dominant poles can be allowed to have different characteristics e.g. all four complex conjugate poles, all four real poles, two real and two complex conjugate poles which have been described in the following sub-sections along with the respective derivations for the dominant pole placement tuning.

### 2.1. *Deriving unique and alternative expressions for stabilizing PID controller gains and obtaining the stability regions*

Using the coefficient matching of the desired *vs.* the given characteristic equation of the closed loop system, we show here that for a chosen set of process parameters $\{K, L, T, \zeta_{ol}\}$, the expressions for integro-differential gains $K_i$ and $K_d$ produce unique values. But there can be many alternative expressions for the proportional gain $K_p$ which can be derived from the two controller gains $K_i$ and $K_d$. The multiple expressions for $K_p$ arise due to an overdetermined system of algebraic equations from various higher powers of the Laplace variable *s*. Therefore, for a given set of plant parameters and design specifications, after obtaining the controller gains $\{K_i, K_d\}$, there can be many possible solutions for $K_p$. The most robust expression for selecting $K_p$ can be found out by comparing the stabilizing regions for all these expressions having the highest volume, compared to the same with the others. This is determined by drawing uniformly distributed random samples from a chosen design parameter space with a specified range and then evaluating the stability at each of these sampled points, as determined by the real part of the closed loop poles being negative.

In order to show this, we use Monte Carlo simulations by sampling the design specifications $\{m, \zeta_{cl}, \omega_{cl}\}$ from chosen intervals for fixed open loop plant parameters $\{\zeta_{ol}, \omega_{ol}, L\}$. The effect of system's DC-gain *K* has not been investigated here, since it simply decreases all the three PID controller gains in a linear fashion. However, the effect of other parameters may be complicated and therefore needs to be explored rigorously using Monte Carlo simulations which is adopted next.

Let us now assume that for a given set of open loop plant parameters $\{\zeta_{ol}, \omega_{ol}, L\}$, one can choose different design specifications $\{m, \zeta_{cl}, \omega_{cl}\}$ to derive the PID controller gains. However, due to the finite term approximation of the time delay, the resulting closed loop system is not necessarily stable for any randomly chosen design parameter or arbitrary PID controller gains within a specified range. We substantiate this argument, using a computational approach of randomly sampling the design specification space within a chosen bound, for fixed open loop SOPTD process parameters which yields a set of stabilizing PID controller gains $\{K_p, K_i, K_d\}$. However, with different design specifications $\{m, \zeta_{cl}, \omega_{cl}\}$, the closed loop system may be stable, but the closed loop performance is expected to vary widely within this entire stability region. Such thousands of randomly sampled guess values for the design specifications yielding stable closed loop systems can be quantified as a fraction of the accepted stable samples to the total number of samples drawn from the entire design specification space. For a more robust system design, the method producing largest stabilizing volume in the design parameter space or the equivalent controller parameter space should be selected.

For the set of stabilizing controller gains, we also calculate various closed loop performance measures like set-point tracking, disturbance rejection, control effort to investigate their trade-offs (Das & Pan 2014), peak sensitivity and complementary sensitivity functions etc. (Åström & Hägglund 2004) signifying parametric robustness and high frequency measurement noise rejection performances respectively. We also demonstrate here the wide applicability of our proposed approach on three different class of SOPTD processes e.g. lag-dominant ($T > L$), delay dominant ($T < L$) and balanced lag-delay ($T \approx L$) system, also each of them with three different damping levels *viz.* overdamped ($\zeta_{ol} > 1$), critically damped ($\zeta_{ol} = 1$), and underdamped ($\zeta_{ol} < 1$). Therefore, we have nine different cases to explore, as shown in the subsequent sections. For the Monte Carlo simulations, the ranges for the desired control performance parameters are chosen as $m \in [1,10], \zeta_{cl} = [1,5], \omega_{cl} = [1,10]$, as per the previous reports like (Panda et al. 2004)(Wang et al. 2009).

## 2.2. *Pole placement PID controller design with all non-dominant complex conjugate poles*

Now, as discussed in (Kiong et al. 1999)(Åström & Hägglund 1995) in order to ensure guaranteed dominant pole placement with PID controllers, let us consider that the closed loop system (5) has six complex (conjugate) poles. Amongst these six, two of them are dominant meeting the desired closed loop design specifications $\{\zeta_{cl}, \omega_{cl}\}$. The rest four poles are non-dominant in nature and their locations can be controlled by selecting the non-dominance pole placement parameter *m*. One can easily choose the *m* in such a way that these poles do not have any significant effect on the closed loop performance of the process with design specifications $\{\zeta_{cl}, \omega_{cl}\}$. Under this assumption, the non-dominant closed loop complex conjugate poles become $s_{1,2}^{nd} = m\left(-\zeta_{cl}\omega_{cl} \pm j\omega_{cl}\sqrt{1-\zeta_{cl}^2}\right)$. Therefore, the resulting characteristic polynomial with both dominant and non-dominant poles can be written as:

$$\begin{aligned}
\Delta(s) &= \left(s^2 + 2\zeta_{cl}\omega_{cl}s + \omega_{cl}^2\right)\left(s - s^{nd}\right)^4 \\
&= \left(s^2 + 2\zeta_{cl}\omega_{cl}s + \omega_{cl}^2\right)\left(\left(s - s_1^{nd}\right)\left(s - s_2^{nd}\right)\right)^2 \\
&= \left(s^2 + 2\zeta_{cl}\omega_{cl}s + \omega_{cl}^2\right)\left(s^2 - \left(s_1^{nd} + s_2^{nd}\right) + s_1^{nd}s_2^{nd}\right)^2 \\
&= \left(s^2 + 2\zeta_{cl}\omega_{cl}s + \omega_{cl}^2\right)\left(s^2 + 2m\zeta_{cl}\omega_{cl}s + \left(m^2\zeta_{cl}^2\omega_{cl}^2 + m^2\omega_{cl}^2\left(1-\zeta_{cl}^2\right)\right)\right)^2 \\
&= \left(s^2 + 2\zeta_{cl}\omega_{cl}s + \omega_{cl}^2\right)\left(s^2 + 2m\zeta_{cl}\omega_{cl}s + m^2\omega_{cl}^2\right)^2 = 0.
\end{aligned} \quad (6)$$

In order to find out the controller parameters using the co-efficient matching method which is also reported in (Kiong et al. 1999)(Åström & Hägglund 1995) for the delay free cases, the closed loop characteristic polynomial (6) can be expanded in terms of the open loop process parameters and the PID controller gains as:

$$\begin{aligned}
&s^6 \left[ L^3 \right] \\
&+ s^5 \left[ 12L^2 + 2\zeta_{ol}\omega_{ol}L^3 - KK_d L^3 \right] \\
&+ s^4 \left[ 60L + 24\zeta_{ol}\omega_{ol}L^2 + \omega_{ol}^2 L^3 + 12KK_d L^2 - KK_p L^3 \right] \\
&+ s^3 \left[ 120 + 120\zeta_{ol}\omega_{ol}L + 12\omega_{ol}^2 L^2 - 60KK_d L + 12KK_p L^2 - KK_i L^3 \right] \\
&+ s^2 \left[ 240\zeta_{ol}\omega_{ol} + 60\omega_{ol}^2 L + 120KK_d - 60KK_p L + 12KK_i L^2 \right] \\
&+ s^1 \left[ 120\omega_{ol}^2 + 120KK_p - 60KK_i L \right] \\
&+ s^0 \left[ 120KK_i \right] \\
&= 0.
\end{aligned} \quad (7)$$

Now the characteristic polynomial in terms of the desired closed loop poles in (6) can also be written as:

$$\begin{aligned}
&s^6 [1] \\
&+ s^5 \left[ 4m\zeta_{cl}\omega_{cl} + 2\zeta_{cl}\omega_{cl} \right] \\
&+ s^4 \left[ 2m^2\omega_{cl}^2 \left(1 + 2\zeta_{cl}^2\right) + 8m\zeta_{cl}^2\omega_{cl}^2 + \omega_{cl}^2 \right] \\
&+ s^3 \left[ 4m^3\zeta_{cl}\omega_{cl}^3 + 4m^2\zeta_{cl}\omega_{cl}^3 \left(1 + 2\zeta_{cl}^2\right) + 4m\zeta_{cl}\omega_{cl}^3 \right] \\
&+ s^2 \left[ m^4\omega_{cl}^4 + 8m^3\zeta_{cl}^2\omega_{cl}^4 + 2m^2\omega_{cl}^4 \left(1 + 2\zeta_{cl}^2\right) \right] \\
&+ s^1 \left[ 2m^4\zeta_{cl}\omega_{cl}^5 + 4m^3\zeta_{cl}\omega_{cl}^5 \right] \\
&+ s^0 \left[ m^4\omega_{cl}^6 \right] \\
&= 0.
\end{aligned} \quad (8)$$

For comparing the coefficients of these two characteristic polynomials, after dividing (7) by $L^3$ and comparing it with (8), the corresponding PID controller gains can be easily obtained as:

$$\begin{aligned}
s^0: \quad & K_i = \frac{m^4 \omega_{cl}^6 L^3}{120K} \\
s^5: \quad & K_d = \left[ \frac{12}{L} + 2\zeta_{ol}\omega_{ol} - 2\zeta_{cl}\omega_{cl}(1 + 2m) \right] / K \\
s^1: \quad & K_p = \left[ 4m^3\zeta_{cl}\omega_{cl}^5 L^3 (m+2) + m^4\omega_{cl}^6 L^4 - 240\omega_{ol}^2 \right] / 240K \\
s^2: \quad & K_p = L^2 \left[ \frac{240\zeta_{ol}\omega_{ol}}{L^3} + \frac{60\omega_{ol}^2}{L^2} + \frac{120KK_d}{L^3} + \frac{12KK_i}{L} - m^4\omega_{cl}^4 - 8m^3\zeta_{cl}^2\omega_{cl}^4 - 2m^2\omega_{cl}^4(1 + 2\zeta_{cl}^2) \right] / 60K \\
s^3: \quad & K_p = \left[ L^3 \left( 4m^3\zeta_{cl}\omega_{cl}^3 + 4m^2\zeta_{cl}\omega_{cl}^3(1+2\zeta_{cl}^2) + 4m\zeta_{cl}\omega_{cl}^3 \right) - \left( 120 + 120\zeta_{ol}\omega_{ol}L + 12\omega_{ol}^2 L^2 - 60KK_d L - KK_i L^3 \right) \right] / 12KL^2 \\
s^4: \quad & K_p = \left[ \frac{60}{L^2} + \frac{24\zeta_{ol}\omega_{ol}}{L} + \omega_{ol}^2 + \frac{12KK_d}{L} - 2m^2\omega_{cl}^2(1+2\zeta_{cl}^2) - 8m\zeta_{cl}^2\omega_{cl}^2 - \omega_{cl}^2 \right] / K.
\end{aligned} \quad (9)$$

It is observed from (9) that the expressions for the two PID controller gains $\{K_i, K_d\}$ are unique but the proportional gain $K_p$ can have four possible expressions.

### 2.3. *Pole placement PID controller design with all non-dominant real poles*

Similar to the above closed-loop characteristic equation in order to find the PID controller gains, the non-dominant poles can also be considered as real (instead of imaginary) and still be manipulated by the non-dominance pole placement parameter $m$ while meeting the same dominant closed loop specification $\{\zeta_{cl},\omega_{cl}\}$. Therefore, for all real non-dominant poles, the characteristic equation becomes:

$$\Delta(s) = \left(s^2 + 2\zeta_{cl}\omega_{cl}s + \omega_{cl}^2\right)\left(s + m\zeta_{cl}\omega_{cl}\right)^4 \\ = \left(s^2 + 2\zeta_{cl}\omega_{cl}s + \omega_{cl}^2\right)\left(s^4 + 4m\zeta_{cl}\omega_{cl}s^3 + 6m^2\zeta_{cl}^2\omega_{cl}^2s^2 + 4m^3\zeta_{cl}^3\omega_{cl}^3s + m^4\zeta_{cl}^4\omega_{cl}^4\right) = 0. \quad (10)$$

The desired characteristic polynomial (10) can be expanded and rearranged according to the power of Laplace variable $s$, as required for the coefficient matching with that of the open loop system with PID controller in (7), which yields:

$$s^6[1] \\ +s^5\left[4m\zeta_{cl}\omega_{cl} + 2\zeta_{cl}\omega_{cl}\right] \\ +s^4\left[6m^2\zeta_{cl}^2\omega_{cl}^2 + 8m\zeta_{cl}^2\omega_{cl}^2 + \omega_{cl}^2\right] \\ +s^3\left[4m^3\zeta_{cl}^3\omega_{cl}^3 + 12m^2\zeta_{cl}^3\omega_{cl}^3 + 4m\zeta_{cl}\omega_{cl}^3\right] \\ +s^2\left[m^4\zeta_{cl}^4\omega_{cl}^4 + 8m^3\zeta_{cl}^4\omega_{cl}^4 + 6m^2\zeta_{cl}^2\omega_{cl}^4\right] \\ +s^1\left[2m^4\zeta_{cl}^5\omega_{cl}^5 + 4m^3\zeta_{cl}^3\omega_{cl}^5\right] \\ +s^0\left[m^4\zeta_{cl}^4\omega_{cl}^6\right] \\ = 0. \quad (11)$$

Now in a similar way dividing (7) by $L^3$ and matching the coefficients of $s$ with (11) yields the corresponding PID controller gains for the real non-dominant poles as:

$$\begin{aligned}
s^0: & \quad K_i = m^4\zeta_{cl}^4\omega_{cl}^6 L^3 / 120K \\
s^5: & \quad K_d = \left[\frac{12}{L} + 2\zeta_{ol}\omega_{ol} - 2\zeta_{cl}\omega_{cl}(1-2m)\right] / K \\
s^1: & \quad K_p = \left[L^3\left(2m^4\zeta_{cl}^5\omega_{cl}^5 + 4m^3\zeta_{cl}^3\omega_{cl}^5\right) - 120\omega_{ol}^2 + 60KK_iL\right] / 120K \\
s^2: & \quad K_p = L^2\left[\frac{240\zeta_{ol}\omega_{ol}}{L^3} + \frac{60\omega_{ol}^2}{L^2} + \frac{120KK_d}{L^3} + \frac{12KK_i}{L} - m^4\zeta_{cl}^4\omega_{cl}^4 - 8m^3\zeta_{cl}^4\omega_{cl}^4 - 6m^2\zeta_{cl}^2\omega_{cl}^4\right] / 60K \\
s^3: & \quad K_p = \left[L^3\left(4m^3\zeta_{cl}^3\omega_{cl}^3 + 12m^2\zeta_{cl}^3\omega_{cl}^3 + 4m^2\zeta_{cl}^2\omega_{cl}^4\right) - \left(120 + 120\zeta_{ol}\omega_{ol}L + 12\omega_{ol}^2L^2 - 60KK_dL - KK_iL^3\right)\right] / 12KL^2 \\
s^4: & \quad K_p = \left[\frac{60}{L^2} + \frac{24\zeta_{ol}\omega_{ol}}{L} + \omega_{ol}^2 + \frac{12KK_d}{L} - 6m^2\zeta_{cl}^2\omega_{cl}^2 - 8m\zeta_{cl}^2\omega_{cl}^2 - \omega_{cl}^2\right] / K.
\end{aligned} \quad (12)$$

Here also, the two PID controller gains $\{K_i, K_d\}$ can be uniquely derived from the open loop and desired process parameters, but the proportional gain $K_p$ can have four different expressions.

### 2.4. *Pole placement PID controller design with two real and two complex conjugate non-dominant poles*

The characteristic equation using two real and two complex conjugate non-dominant poles can be written as:

$$\Delta(s) = \left(s^2 + 2\zeta_{cl}\omega_{cl}s + \omega_{cl}^2\right)\left(s^2 + 2m\zeta_{cl}\omega_{cl}s + m^2\omega_{cl}^2\right)\left(s + m\zeta_{cl}\omega_{cl}\right)^2 = 0. \quad (13)$$

This yields the coefficients in decreasing order of Laplace variable $s$ as:

$$\begin{aligned}
&s^6[1]\\
&+s^5\left[4m\zeta_{cl}\omega_{cl}+2\zeta_{cl}\omega_{cl}\right]\\
&+s^4\left[5m^2\zeta_{cl}^2\omega_{cl}^2+m^2\omega_{cl}^2+8m\zeta_{cl}^2\omega_{cl}^2+\omega_{cl}^4\right]\\
&+s^3\left[2m^3\zeta_{cl}^3\omega_{cl}^3+2m^3\zeta_{cl}\omega_{cl}^3+2m^2\zeta_{cl}^3\omega_{cl}^3+8m^2\zeta_{cl}^3\omega_{cl}^3+2m^2\zeta_{cl}\omega_{cl}^3+4m\zeta_{cl}\omega_{cl}^3\right]\\
&+s^2\left[m^4\zeta_{cl}^2\omega_{cl}^4+4m^3\zeta_{cl}^4\omega_{cl}^4+4m^3\zeta_{cl}^2\omega_{cl}^4+m^2\zeta_{cl}^2\omega_{cl}^4+4m^2\zeta_{cl}^2\omega_{cl}^4+m^2\omega_{cl}^4\right]\\
&+s^1\left[2m^4\zeta_{cl}^3\omega_{cl}^5+2m^3\zeta_{cl}^3\omega_{cl}^5+2m^3\zeta_{cl}\omega_{cl}^5\right]\\
&+s^0\left[m^4\zeta_{cl}^2\omega_{cl}^6\right]\\
&=0.
\end{aligned} \qquad (14)$$

Again, dividing (7) by $L^3$ and match the co-efficient with (14) as done previously, yields the corresponding PID controller gains as:

$$\begin{aligned}
s^0: &\ K_i = m^4\zeta_{cl}^2\omega_{cl}^6 L^3 / 120K\\
s^5: &\ K_d = \left[\frac{12}{L}+2\zeta_{ol}\omega_{ol}-2\zeta_{cl}\omega_{cl}(1+2m)\right]\Big/K\\
s^1: &\ K_p = \left[L^3\left(2m^4\zeta_{cl}^3\omega_{cl}^5+2m^3\zeta_{cl}^3\omega_{cl}^5+2m^3\zeta_{cl}\omega_{cl}^5\right)-120\omega_{ol}^2+60KK_iL\right]\Big/120K\\
s^2: &\ K_p = L^2\left[\frac{240\zeta_{ol}\omega_{ol}}{L^3}+\frac{60\omega_{ol}^2}{L^2}+\frac{120KK_d}{L^3}+\frac{12KK_i}{L}-\left(m^4\zeta_{cl}^2\omega_{cl}^4+4m^3\zeta_{cl}^4\omega_{cl}^4+4m^3\zeta_{cl}^2\omega_{cl}^4+m^2\zeta_{cl}^2\omega_{cl}^4+4m^2\zeta_{cl}^2\omega_{cl}^4+m^2\omega_{cl}^4\right)\right]\Big/60K\\
s^3: &\ K_p = \left[\begin{array}{l}L^3\left(2m^3\zeta_{cl}^3\omega_{cl}^3+2m^3\zeta_{cl}\omega_{cl}^3+2m^2\zeta_{cl}^3\omega_{cl}^3+8m^2\zeta_{cl}^3\omega_{cl}^3+2m^2\zeta_{cl}\omega_{cl}^3+4m\zeta_{cl}\omega_{cl}^3\right)\\-\left(120+120\zeta_{ol}\omega_{ol}L+12\omega_{ol}^2L^2-60KK_dL-KK_iL^3\right)\end{array}\right]\Big/12KL^2\\
s^4: &\ K_p = \left[\frac{60}{L^2}+\frac{24\zeta_{ol}\omega_{ol}}{L}+\omega_{ol}^2+\frac{12KK_d}{L}-\left(5m^2\zeta_{cl}^2\omega_{cl}^2+m^2\omega_{cl}^2+8m\zeta_{cl}^2\omega_{cl}^2+\omega_{cl}^4\right)\right]\Big/K.
\end{aligned} \qquad (15)$$

Similar to the previous cases, $\{K_i, K_d\}$ gains have unique expressions but the gain $K_p$ has four possible values depending on the choice of different coefficients of Laplace variable $s$.

### 3. Designing robust stable PID controller and evaluation of different closed loop performance measures

#### 3.1. Determining the robust stable solutions using centroid of the stability region

We here use the *k*-means clustering algorithm to determine the centroid of the stability region in the PID controller space which can have a complex shape in the 3D parameter space of the controller gains $\{K_p, K_i, K_d\}$. The stability regions are determined for nine classes of SOPTD plants and it is also checked that a single centroid represents the stability regions of a unimodal distribution in the controller parameter space. Otherwise if a multi-modal distribution is discovered, indicating more than one possible robust stable solution for the controller gains, the number of centroids (*k*) in the *k*-means clustering algorithm can also be set to the number of modes in the distribution of controller gains. Previously determination of the stability region centroids in the controller parameter space were done by looking for only circular (for two gains), spherical (for three gains) or (hyper-)spherical clusters (more than two controller gains) as shown in (Pan et al. 2011), which has been extended here to find out centroids of more generalised complicated shaped clusters like ellipsoids or other structures by assuming that the centroid is likely to lie in the high density region of the clusters where most of the samples are accepted in the random Monte Carlo sampling.

Although the chosen design parameter space has been explored using $10^5$ uniformly distributed random samples, certain areas of the design specification or the equivalent controller parameter space has more stable solutions than the others. Therefore, the present random sampling approach helps in understanding the shape of the stability region using various expressions for deriving the PID controller gains, while also giving the flexibility to specify the nature of the non-dominant poles (i.e. all real, all complex or mixed), as explored in the previous sections.

The *k*-means clustering algorithm starts with a random initial guess for the centroid of the multi-dimensional data space and iteratively move the centroids based on minimizing the squared Euclidean distance criteria from all the data points (Rogers & Girolami 2012). However, to ensure that the best possible estimate of the centroid or the robust stable solution has been discovered in the iterative process, the *k*-means clustering has been run 10 times, for each case with random starting points and the best solution with minimum Euclidean distance is chosen as the final estimate of the centroid, signifying the most robust stable solution in the PID controller parameter space.

Therefore, for a given process model characterised by the constants $\{K, \zeta_{ol}, \omega_{ol}, L\}$, the robust stable PID controller gains obtained *via* the dominant pole placement can be obtained using the following steps:

***Step 1:*** Choose the non-dominant pole types amongst all complex, all real or mixed using the expressions in (9), (12) or (15) to map the open loop and desired closed loop system parameters on to the PID controller gains.

***Step 2:*** Obtain the stabilizing PID controller gains using few thousands of uniformly distributed samples within a chosen range of all design parameters $\{m, \zeta_{cl}, \omega_{cl}\}$.

***Step 3:*** Cluster the stabilizing PID gains to get the centroid as the robust stable solution.

***Step 4***: Evaluate different performance measures with the robust stable PID controller on the nominal and perturbed process models.

### 3.2. *Performance measures with the stabilizing PID controller gains*

The previous section derives the expressions for obtaining stabilizing PID controller gains by choosing the nature of the non-dominant poles being all complex, real or mixture of them. However, in a realistic control system design problem, apart from the stability, the control loop performance is also of major concern. Therefore, the robust stable solutions might not always show an acceptable control performance. On contrary the controller setting for optimised performance criteria may not have a sufficient robustness against plant or controller parameter variation. Therefore, we chose to design the PID controller settings based on maximum robust stability and then compare the achievable control performances. However the robustness checking of optimally designed controllers is also a valid approach as previously studied for single control objective (Das, Pan, et al. 2012; Das et al. 2011) and multiple conflicting control objectives (Das et al. 2015)(Das & Pan 2014). Also, the parametric robustness of optimal controllers and optimality of robust design have been previously discussed in (Pan & Das 2016).

After the robust stable solution is determined using clustering which yields the centroid of the arbitrary shaped stability region, few well-known performance measures are calculated next to compare the effect of having different type of non-dominant poles. Amongst these, both time and frequency domain performance measures are evaluated, however both of them are neither specified nor can be guaranteed together, under the present design approach, as also shown in (Das et al. 2011). However for a fair comparison we evaluate different control performances with the PID controller gains obtained using different non-dominant pole types e.g. gain and phase margin ($G_m$ and $\Phi_m$) and gain crossover frequency ($\omega_{gc}$) controlling the overshoot *vs.* speed of operation (Das et al. 2011), peak sensitivity and complementary sensitivity ($M_s$ and $M_T$) (Åström & Hägglund 2004), $\mathcal{H}_2/\mathcal{H}_\infty$ norms for tracking or command following mode (servo) and disturbance rejection (regulatory) mode (Alcántara et al. 2013)(Arrieta et al. 2010).

Let, $G_{ol}(s)$ be the open-loop transfer function comprising of the time-delayed SOPTD system $G(s)$ with PID controller $C(s)$ which can be represented as:

$$G_{ol}(s) = C(s)G(s). \tag{16}$$

The basic PID control loop with different inputs e.g. set-point ($r$), disturbance ($d$) and noise ($n$) and the measurement points e.g. error ($x_1$), control signal ($x_2$) and noisy process variable ($x_3$) are shown in Figure 1. To guarantee internal stability and also for evaluating different performance measures of a feedback control loop the following nine transfer functions in (17) play a major role (Doyle et al. 2013)

$$\begin{bmatrix} x_1 \\ x_2 \\ x_3 \end{bmatrix} = \begin{bmatrix} e \\ u+d \\ y+n \end{bmatrix} = \frac{1}{1+GC} \begin{bmatrix} 1 & -G & -1 \\ C & 1 & -C \\ GC & G & 1 \end{bmatrix} \begin{bmatrix} r \\ d \\ n \end{bmatrix}. \tag{17}$$

These matrix shows the effective transfer functions from different inputs to various measurement points. Amongst the nine, four transfer functions play the central role to characterise the control system performance (Doyle et al. 2013)(Das & Pan 2014)(Herreros et al. 2002), i.e. the sensitivity $S^e(s)$, complementary sensitivity $T(s)$, disturbance sensitivity $S^d(s)$ and control sensitivity $S^u(s)$ as follows:

$$S^e(s) = \frac{1}{1+G_{ol}(s)} = G_{re}(s) = G_{dx_2}(s) = G_{nx_3}(s),$$

$$T(s) = \frac{G_{ol}(s)}{1+G_{ol}(s)} = G_{rx_3}(s), \quad S^e(s) + T(s) = 1, \tag{18}$$

$$S^d(s) = \frac{G(s)}{1+G_{ol}(s)} = G_{dx_3}(s), \quad S^u(s) = \frac{C(s)}{1+G_{ol}(s)} = G_{rx_2}(s).$$

These four transfer functions can be quantified using various systems norms ($\mathcal{H}_2/\mathcal{H}_\infty$) for different system inputs (Herreros et al. 2002). It is clear from (18) that the sensitivity function $S^e(s)$ has three different interpretations. In other words, the sensitivity function signifies the effective transfer function at all the three measurement points in Figure 1, for set-point, disturbance and noise inputs, as also revealed from the same diagonal elements in (17). Therefore, in a process control design, it has been considered as one of the fundamentally important criteria which is often considered to be norm bounded in various tuning rules like MIGO (M-constrained integral gain optimization) and approximate MIGO (AMIGO) etc. for PI/PID controller design (Åström & Hägglund 2004; Hägglund & Åström 2004)(Hägglund & Åström 2002).

First we check the performance $(J_d)$ for a step disturbance input $(d(s))$ and calculating the $\mathcal{H}_2/\mathcal{H}_\infty$ norms of the disturbance sensitivity function which can be represented by:

$$\begin{aligned} J_2^d &= \|d_1(s)S_d(s)\|_2, d_1(s) = 1/s, \\ J_\infty^d &= \|d_2(s)S_d(s)\|_\infty, d_2(s) = 1/s. \end{aligned} \tag{19}$$

Using the final value theorem for Laplace transform, the functions with finite time domain integral are quantified and accordingly the set-point/disturbance inputs are selected between step/impulse as $\{r,d\}=1/s$ or $\{r,d\}=1$ respectively, as the input to different sensitivity functions.

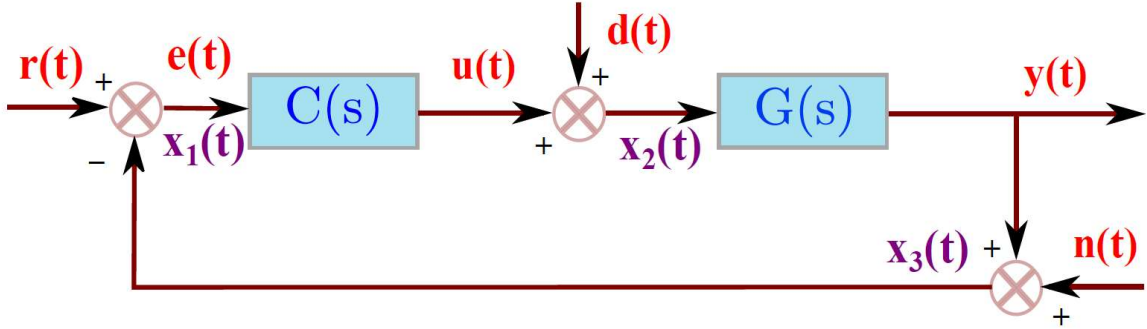

Figure 1: Schematic diagram of the PID control loop with different inputs and measurement points.

The controller output or the control signal leads to the actuator limits and frequent oscillatory inputs to the actuator $J^u$ which can be represented by the $\mathcal{H}_2/\mathcal{H}_\infty$ norms of the control sensitivity function. However, for standard PID controller structure without a derivative filter as being used here for obtaining 3 controller parameter based 3D stability regions, the control sensitivity becomes an improper transfer function with more zeros than poles which forbids calculating system norms directly. As an alternative approach, the $\mathcal{L}_2/\mathcal{L}_\infty$ norms of the control signal can be computed for a step input to $S^u(s)$ if it is proper, or impulse input to $S^u(s)/s$ if the control sensitivity is improper, as in the present case. Therefore the control sensitivity norms can be quantified as the $\mathcal{L}_2/\mathcal{L}_\infty$ norm for step change in the set-point using inverse-Laplace transform ($\mathbb{L}^{-1}$) of $S^u(s)/s$ as:

$$\begin{aligned}J_2^u &= \|S^u(t)\|_2 = \|\mathbb{L}^{-1}(d_1(s)S^u(s))\|_2, d_1(s)=1/s,\\ J_\infty^u &= \|S^u(t)\|_\infty = \|\mathbb{L}^{-1}(d_2(s)S^u(s))\|_\infty, d_2(s)=1/s.\end{aligned} \quad (20)$$

The set point tracking performance $J^e$ is analysed using the sensitivity function subjected to step change in the set-point yielding

$$\begin{aligned}J_2^e &= \|d_1(s)S^e(s)\|_2, d_1(s)=1/s,\\ J_\infty^e &= \|d_2(s)S^e(s)\|_\infty, d_2(s)=1/s.\end{aligned} \quad (21)$$

The noise rejection performance can be quantified as $J^n$ using the complementary sensitivity function subjected to impulse input in the set-point yielding

$$\begin{aligned}J_2^n &= \|d_1(s)T(s)\|_2, d_1(s)=1,\\ J_\infty^n &= \|d_2(s)T(s)\|_\infty, d_2(s)=1.\end{aligned} \quad (22)$$

For all the different sensitivity functions, the corresponding $\mathcal{H}_2/\mathcal{H}_\infty$ system norms and $\mathcal{L}_2/\mathcal{L}_\infty$ signal norms are defined as:

$$\|G(s)\|_2 = \sqrt{\frac{1}{2\pi}\int_{-\infty}^{\infty}|G(j\omega)|^2 d\omega}, \quad \|G(s)\|_\infty = \sup_\omega |G(j\omega)|,$$

$$\|g(t)\|_2 = \sqrt{\int_{-\infty}^{\infty}u(t)^2 dt}, \quad \|g(t)\|_\infty = \sup_t |g(t)|. \tag{23}$$

The $\mathcal{H}_2$ norms here in (19), (21) and (22) represent large and sustained oscillations in the disturbance response, error signal and process variable, whereas the $\mathcal{H}_\infty$ norms denote the peak gain of the frequency response for a particular type of input (servo/regulatory). For stable closed loop system, the corresponding $\mathcal{H}_2$ norms of the sensitivity functions with a chosen type of input excitation need to be finite and should have a low-pass characteristic i.e. as $\omega \to \infty$, the Bode magnitude plot should be drooping in nature.

Next we also compute the gain margin ($G_m$), phase margin ($\Phi_m$) and the gain cross-over frequency ($\omega_{gc}$) of the open loop system which signify the robustness, oscillatory nature and speed of the closed loop response respectively. These three important quantities for a given process model and set of PID controller gains can be derived as:

$$\begin{aligned}
Arg\left[G_{ol}(j\omega_{gc})\right] &= Arg\left[C(j\omega_{gc})G(j\omega_{gc})\right] = -\pi + \Phi_m \\
\left|G_{ol}(j\omega_{gc})\right| &= \left|C(j\omega_{gc})G(j\omega_{gc})\right| = 1 \\
\left|G_{ol}(j\omega_{pc})\right| &= \left|C(j\omega_{pc})G(j\omega_{pc})\right| = 1/G_m \\
Arg\left[G_{ol}(j\omega_{pc})\right] &= Arg\left[C(j\omega_{pc})G(j\omega_{pc})\right] = -\pi.
\end{aligned} \tag{24}$$

## 4. Simulation and results
### 4.1. Test-bench SOPTD processes for performance evaluation

For each of the nine classes of test-bench plants under investigation, we also tabulate which pole configuration and expression for the PID controller gains yield the largest stability region, as explored from the number of stabilizing solutions obtained from the Monte Carlo simulations on the chosen design parameter space. The percentage volume can be represented as the ratio of accepted stable solutions to the total number of uniformly distributed ($10^5$) random samples drawn from the chosen parameter space. In addition, the correlation amongst the stabilizing controller gains or the design parameters are also investigated. For example, it can be observed that a high $\omega_{cl}$ can only be achieved with small $m$ i.e. allowing more effect of the non-dominant poles, thus unnecessarily affecting the control performance. Similarly, an inverse relation is observed between the stabilizing $\omega_{cl}$ and $\zeta_{cl}$. In principle, a high value of $m$ is desired which keeps the effect of non-dominant poles on the control performance as minimum. However it will be evident from the simulation examples presented later that the high $m$ regions are sparse and is also limiting the design specification on the $\omega_{cl}$ and $\zeta_{cl}$.

The nine classes of test-bench processes and their characteristics are as follows:

i)  *Lag-dominant (L<T) underdamped ($\zeta_{ol}$<1) process* (Jahanmiri & Fallahi 1997)

$$G_1(s) = \frac{1}{9s^2 + 2.4s + 1}e^{-s}, \tag{25}$$

with $K = 1/9$, $L = 1$, $T = 3$, $\zeta_{ol} = 0.4$, $L/T = 0.333<1$.

ii)  *Lag-dominant (L<T) critically-damped ($\zeta_{ol}$=1) process* (Hwang & others 1995)

$$G_2(s) = \frac{1}{s^2 + 2s + 1} e^{-0.8s}, \tag{26}$$

with $K = 1$, $L = 0.8$, $T = 1$, $\zeta_{ol} = 1$, $L/T = 0.8 < 1$.

    iii)    *Lag-dominant (L<T) overdamped ($\zeta_{ol}$>1) process* (Pomerleau et al. 1996)

$$G_3(s) = \frac{e^{-2s}}{(1+10s)(1+4s)}, \tag{27}$$

with $K = 1/40$, $L = 2$, $T = 6.3246$, $\zeta_{ol} = 1.1068$, $L/T = 0.3162 < 1$.

    iv)    *Balanced lag-delay (L≈T) underdamped ($\zeta_{ol}$<1) process* (Hwang & others 1995)

$$G_4(s) = \frac{0.5}{s^2 + 1.2s + 1} e^{-s}, \tag{28}$$

with $K = 0.5$, $L = 1$, $T = 1$, $\zeta_{ol} = 0.6$, $L/T = 1$.

    v)    *Balanced lag-delay (L≈T) critically-damped ($\zeta_{ol}$=1) process* (Hägglund & Åström 2004)

$$G_5(s) = \frac{e^{-s}}{(1+s)^2}, \tag{29}$$

with $K = 1$, $L = 1$, $T = 1$, $\zeta_{ol} = 1$, $L/T = 1$.

    vi)    *Balanced lag-delay (L≈T) overdamped ($\zeta_{ol}$>1) process* (Panda et al. 2004)

$$G_6(s) = \frac{e^{-3s}}{9s^2 + 24s + 1}, \tag{30}$$

with $K = 1/9$, $L = 3$, $T = 3$, $\zeta_{ol} = 4$, $L/T = 1$.

    vii)    *Delay dominant (L>T) underdamped ($\zeta_{ol}$<1) process* (Wang & Shao 2000)

$$G_7(s) = \frac{e^{-1.2755s}}{3.2158s^2 + 3.1614s + 3.0568}, \tag{31}$$

with $K = 1/3.2158$, $L = 1.2755$, $T = 1.0257$, $\zeta_{ol} = 0.5042$, $L/T = 1.2435 > 1$.
This is a reduced order model of a highly oscillatory higher order process

$$G(s) = \frac{1}{(s^2 + s + 1)(s + 3)} e^{-s}.$$

    viii)    *Delay dominant (L>T) critically-damped ($\zeta_{ol}$=1) process* (Thyagarajan & Yu 2003)

$$G_8(s) = \frac{e^{-10s}}{(s+1)^2}, \tag{32}$$

with $K = 1$, $L = 10$, $T = 1$, $\zeta_{ol} = 1$, $L/T = 10 > 1$.

    ix)    *Delay dominant (L>T) overdamped ($\zeta_{ol}$>1) process* (Bi et al. 2000)

$$G_9(s) = \frac{e^{-2s}}{0.12s^2 + 1.33s + 1.24}, \tag{33}$$

with $K = 1/0.12$, $L = 2$, $T = 0.3111$, $\zeta_{ol} = 1.7239$, $L/T = 6.4288 > 1$. This represents an HVAC system model between fan speed to the supply air pressure control loop.

The next sub-section reports the stability regions of each of these nine classes of test-bench processes, representing different dynamical behaviour with various lag to delay ratio and open loop oscillation levels.

### *4.2. Stability regions and the robust stable PID controller design for the test-bench SOPTD processes*

Out of the 12 possible expressions (3 non-dominant pole types × 4 different coefficient orders of Laplace variable to find out $K_p$) for the stabilizing gains, we now identify the stabilizing data points within a range of design parameters for the expression where the number of stable solutions is maximum. The stabilizing data-points in the design parameter space $\{m, \zeta_{cl}, \omega_{cl}\}$ are next projected on to the controller parameter space $\{K_p, K_i, K_d\}$ by an one to one mapping using the expressions in (9), (12) and (15). These sets of stabilizing controller gains are then fed to a clustering algorithm to find out the centroids of the stabilizing regions, for each of the non-dominant pole types. Next, the performances of these robust stable solutions are also compared, using various criteria introduced in section 3.2.

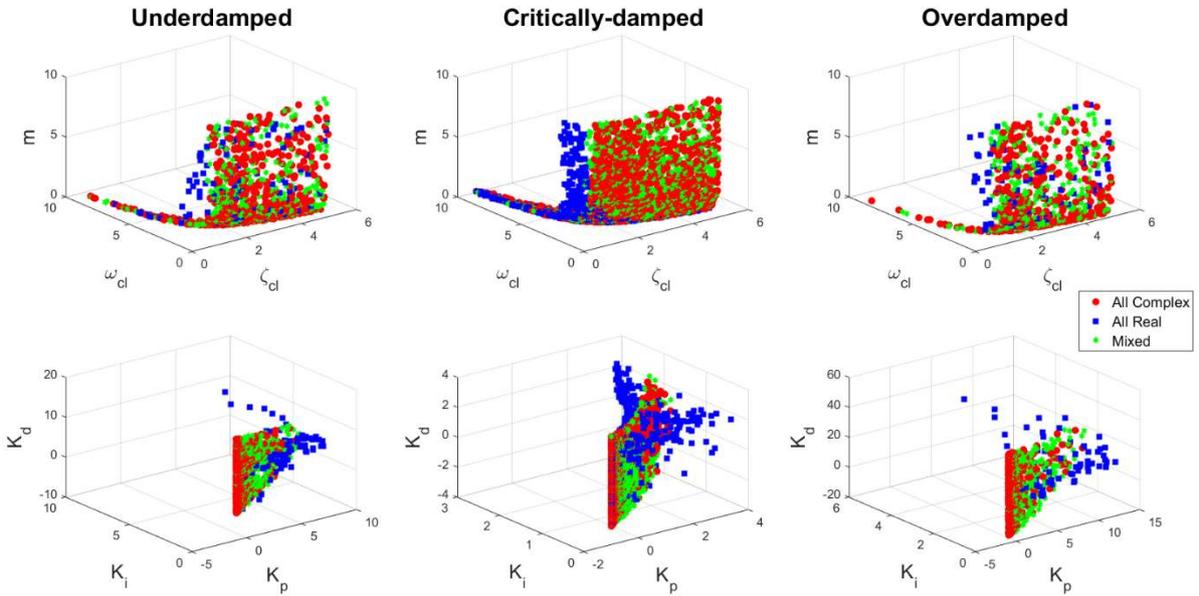

*Figure 2: 3D stability region in the design specification and controller parameter space for the three lag-dominant processes. (top) design parameter space, (bottom) PID controller parameter space.*

For the nine test-bench processes in (25)-(33), the stability regions in the design parameter space (top panels) and PID controller parameter space (bottom panels) are shown in Figure 2-Figure 4 respectively. Out of the $10^5$ uniformly drawn samples in the chosen design parameter space $\{m, \zeta_{cl}, \omega_{cl}\}$, the number of stabilizing solutions are reported in Table 1 for the nine test-bench processes and different non-dominant pole types and expressions for proportional gain $K_p$. It is also evident from Table 1 that the all complex non-dominant pole type and the first ($s^1$) expression for $K_p$, yields a larger stabilizing region given by the percentage (%) volume or the number of stable solutions, obtained from the randomly drawn samples. This is due to the reason that for all four complex non-dominant poles, they have wider flexibility by adjusting their real and imaginary parts to stabilize the closed loop system. However, for the two complex/two real case, lesser number of non-dominant poles can explore a larger portion of the negative *s*-plane. Similarly, a more stringent criterion is imposed for the all real non-dominant poles case, as they are forced to lie only on the negative real axis and cannot explore the entire negative *s*-plane, which gives them lesser degrees of freedom and hence yields smaller stability regions. For each type of non-dominant poles and for all the test-bench processes in Table 1, the expression with highest percentage volume is highlighted in italics. Next considering the robust stable solution as the centroid of the three largest stability regions representing different pole types, we now compare the performances of these robust stable solutions.

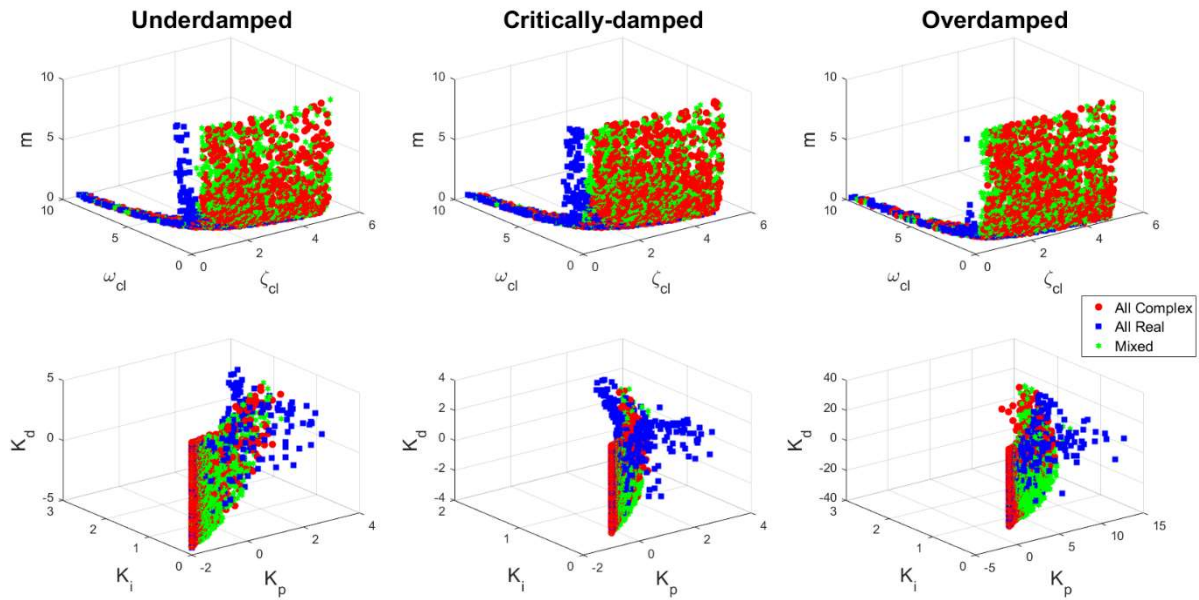

*Figure 3: 3D stability region in the design specification and controller parameter space for the three balanced lag-delay processes. (top) design parameter space, (bottom) PID controller parameter space.*

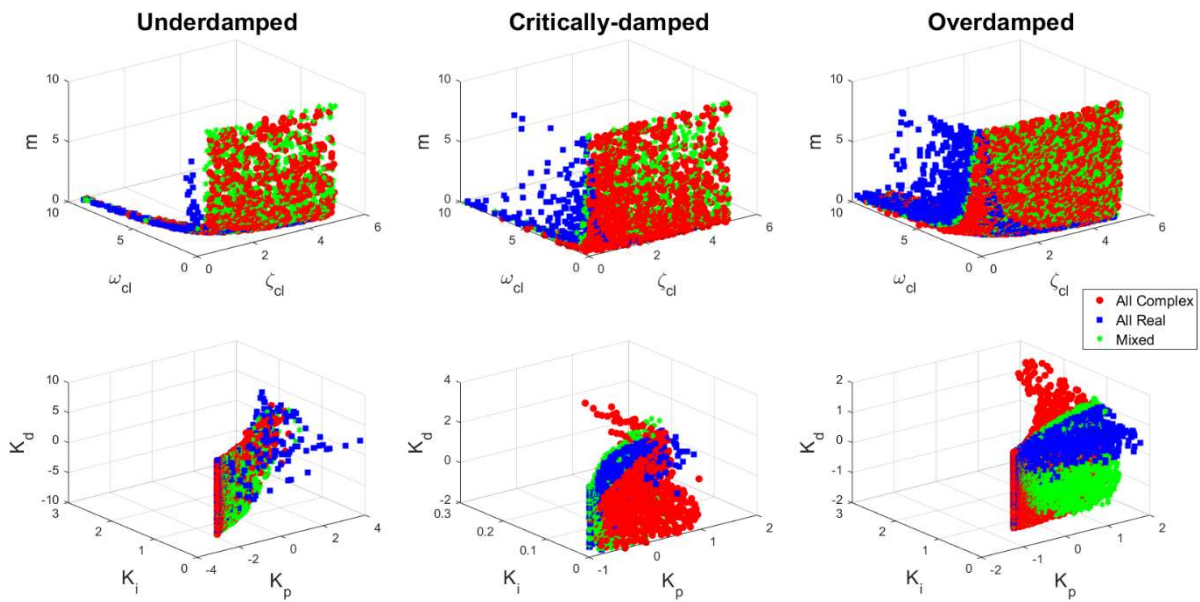

*Figure 4: 3D stability region in the design specification and controller parameter space for the three delay-dominant processes. (top) design parameter space, (bottom) PID controller parameter space.*

It is observed from Figure 2-Figure 4 that the stability regions are wider for high $\zeta_{cl}$ with the two/four complex non-dominant poles, whereas the number of stabilizing data points are only fewer with a demand of high closed loop damping in the case of all real poles. This implies that the centroid of the stabilizing all real non-dominant poles will lie in low closed loop damping region and will yield a less sluggish response than that with the two/four complex non-dominant poles. Also, in the corresponding controller parameter space, the all real non-dominant pole case yields a higher value of $K_d$.

Table 1: Number of stable controller gains for the nine test-bench processes using $10^5$ random Monte Carlo evaluations

| Process type | Open loop damping | Nature of non-dominant pole | Alternate stabilizing expressions for $K_p$ | | | | Max. no. of stable solutions | % Volume |
|---|---|---|---|---|---|---|---|---|
| | | | $s^1$ | $s^2$ | $s^3$ | $s^4$ | | |
| Lag dominant | Underdamped | all complex | *503* | 442 | 33 | 1 | 503 | 0.503 |
| | | all real | 107 | *108* | 32 | 0 | 108 | 0.108 |
| | | two complex/two real | *447* | 39 | 47 | 0 | 447 | 0.447 |
| | Critically-damped | all complex | *1401* | 264 | 242 | 23 | 1401 | 1.401 |
| | | all real | *630* | 330 | 167 | 22 | 630 | 0.63 |
| | | two complex/two real | *1336* | 156 | 140 | 4 | 1336 | 1.336 |
| | Overdamped | all complex | *361* | 292 | 37 | 0 | 361 | 0.361 |
| | | all real | 62 | 11 | *76* | 0 | 76 | 0.076 |
| | | two complex/two real | *351* | 25 | 37 | 0 | 351 | 0.351 |
| Balanced lag and delay | Underdamped | all complex | *1028* | 807 | 142 | 22 | 1028 | 1.028 |
| | | all real | *376* | 316 | 137 | 15 | 376 | 0.376 |
| | | two complex/two real | *1038* | 128 | 108 | 5 | 1038 | 1.038 |
| | Critically-damped | all complex | *1403* | 295 | 237 | 34 | 1403 | 1.403 |
| | | all real | *645* | 423 | 161 | 41 | 645 | 0.645 |
| | | two complex/two real | *1363* | 204 | 140 | 13 | 1363 | 1.363 |
| | Overdamped | all complex | *1356* | 1146 | 268 | 33 | 1356 | 1.356 |
| | | all real | *341* | 32 | 123 | 38 | 341 | 0.341 |
| | | two complex/two real | *1262* | 189 | 142 | 13 | 1262 | 1.262 |
| Delay dominant | Underdamped | all complex | *933* | 766 | 114 | 19 | 933 | 0.933 |
| | | all real | *327* | 199 | 123 | 30 | 327 | 0.327 |
| | | two complex/two real | *933* | 122 | 81 | 6 | 933 | 0.933 |
| | Critically-damped | all complex | 1007 | 365 | *1389* | 180 | 1389 | 1.389 |
| | | all real | *394* | 0 | 384 | 162 | 394 | 0.394 |
| | | two complex/two real | *992* | 313 | 707 | 130 | 992 | 0.992 |
| | Overdamped | all complex | *5043* | 12 | 2487 | 349 | 5043 | 5.043 |
| | | all real | *2462* | 568 | 1057 | 335 | 2462 | 2.462 |
| | | two complex/two real | *4640* | 884 | 2229 | 293 | 4640 | 4.64 |

The stabilizing controller gains having the highest percentage volume amongst the four alternative expressions for $K_p$ in Table 1 are now clustered for each type of test-bench process and non-dominant pole types. The centroids of the stability region as identified by 10 independent runs of the *k*-means clustering algorithm can be considered as the robust stable PID controller gains and are reported in Table 2 for each type of processes and non-dominant pole type. The median distance of all the stabilizing samples from the centroid are also reported in Table 2 indicating compactness of the clusters which is consistently smallest for the all real non-dominant pole type. The projected stability regions in the 2D pair-wise controller parameter space are shown in Figure 5-Figure 13 along with the centroids or the robust stable PID controller gains as the red star, for all the test-bench process types with different lag to delay ratio and open loop oscillation levels. It is evident in most cases that the $K_p$ and $K_i$ (left panels in Figure 5-Figure 13) are more correlated compared to the other pairs of PID controller gains.

We now investigate which cluster has the highest spread in the 3D space of PID controller gains. We also report the median distance between the stabilizing samples and the cluster centroid or the robust stable PID controller gain. The complex shapes of the stabilizing cluster of data-points indicate a highly skewed distribution of the stabilizing controller gains as revealed from their histograms in Figure 14. This skewed distribution is a result of different degree of influence of the design parameters or the equivalent controller gains on the shape of the stability regions for each type of test-bench process. This indicates that a different variation of the PID controller gains is possible along different direction or alternatively different extent of parametric uncertainty in the plant model can be allowed using the robust stable solution for the PID controller gains (Silva et al. 2007). Almost in all the cases of the lag-dominant plants, the all real non-dominant poles have the minimum spread (inter-quartile range) and hence the most compact cluster as revealed from the top panel of Figure 14. However for the delay dominant and balanced lag-delay plants the spread become comparable but creates more outliers indicating distant stabilizing regions far away from the centroid which is also evident from the projected 2D scatter diagrams in Figure 8-Figure 13.

*Table 2: Robust stable gains and median distance of the stabilizing samples from the centroid*

| Process Type | Open loop damping | Nature of non-dominant pole | Median distance | Robust PID gains | | |
|---|---|---|---|---|---|---|
| | | | | $K_p$ | $K_i$ | $K_d$ |
| Lag dominant | Underdamped | all complex | 21.3949 | -0.3412 | 0.0826 | 6.5390 |
| | | all real | 7.8779 | 3.1883 | 0.7876 | 7.5194 |
| | | two complex/two real | 21.1823 | 0.5443 | 0.1902 | 7.4275 |
| | Critically-damped | all complex | 2.6631 | -0.5196 | 0.1229 | 0.5173 |
| | | all real | *3.3457* | 0.6856 | 0.6052 | 1.2991 |
| | | two complex/two real | 2.5681 | -0.2923 | 0.1630 | 0.8199 |
| | Overdamped | all complex | 189.5916 | 0.1429 | 0.0827 | 15.0279 |
| | | all real | 39.5518 | 6.2367 | 0.7632 | 17.0616 |
| | | two complex/two real | 161.5911 | 1.3567 | 0.1650 | 15.6003 |
| Balanced lag and delay | Underdamped | all complex | 5.2558 | -1.4974 | 0.1001 | 0.4146 |
| | | all real | *6.0959* | -0.0842 | 0.5029 | 1.5163 |
| | | two complex/two real | 5.2462 | -1.2407 | 0.1305 | 0.7598 |
| | Critically-damped | all complex | 2.2106 | -0.6477 | 0.0743 | 0.3124 |
| | | all real | *2.5821* | 0.3531 | 0.3623 | 1.0217 |
| | | two complex/two real | 2.2602 | -0.4108 | 0.1177 | 0.4655 |
| | Overdamped | all complex | 134.3780 | 0.4472 | 0.1281 | 3.2445 |
| | | all real | 108.3602 | 2.4796 | 0.2550 | 7.9074 |
| | | two complex/two real | 120.4336 | 1.3153 | 0.1828 | 6.5032 |
| Delay dominant | Underdamped | all complex | 9.2729 | -2.4113 | 0.1106 | 0.1404 |
| | | all real | *10.0374* | -1.0152 | 0.3795 | 1.1933 |
| | | two complex/two real | 9.8668 | -2.1395 | 0.1349 | 0.0955 |
| | Critically-damped | all complex | 1.2044 | 0.0140 | 0.0096 | 1.1905 |
| | | all real | 0.8523 | -0.2052 | 0.0199 | 1.3696 |
| | | two complex/two real | 0.9227 | -0.5277 | 0.0179 | 0.3647 |
| | Overdamped | all complex | 0.6347 | -0.7776 | 0.0962 | 0.2456 |
| | | all real | *0.7960* | -0.2768 | 0.1922 | 0.7399 |
| | | two complex/two real | 0.6297 | -0.5778 | 0.0984 | 0.4143 |

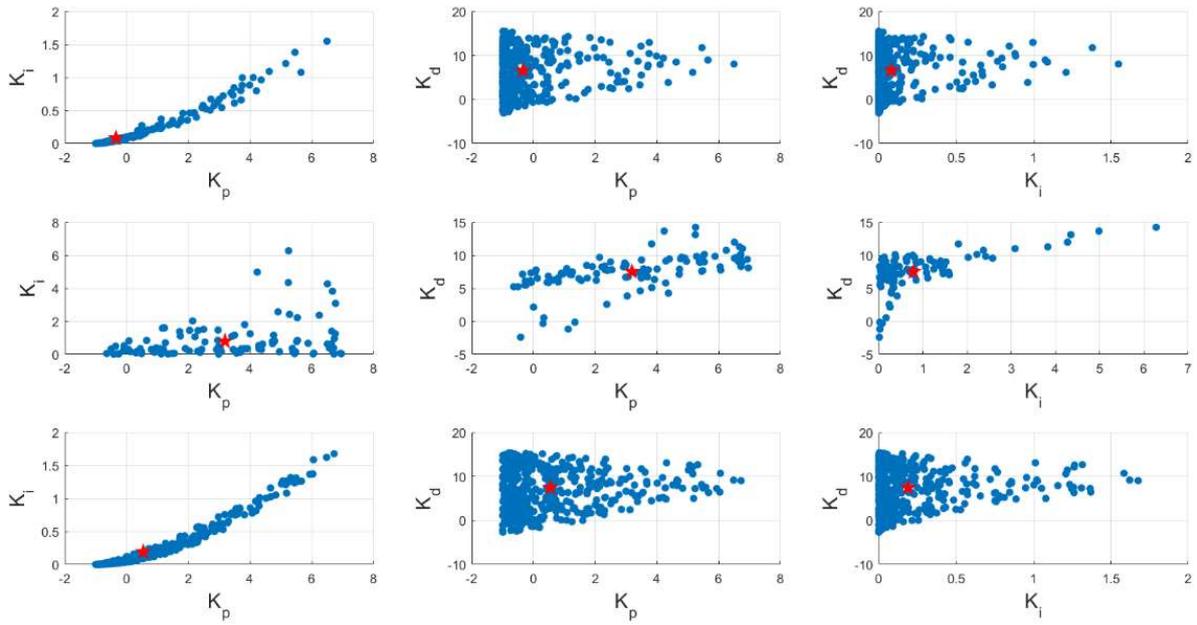

*Figure 5: Stability regions of PID controller gains to control test-bench plant $G_1$ using different non-dominant pole types as (top) all complex poles, (middle) all real poles, (bottom) two real and two complex poles.*

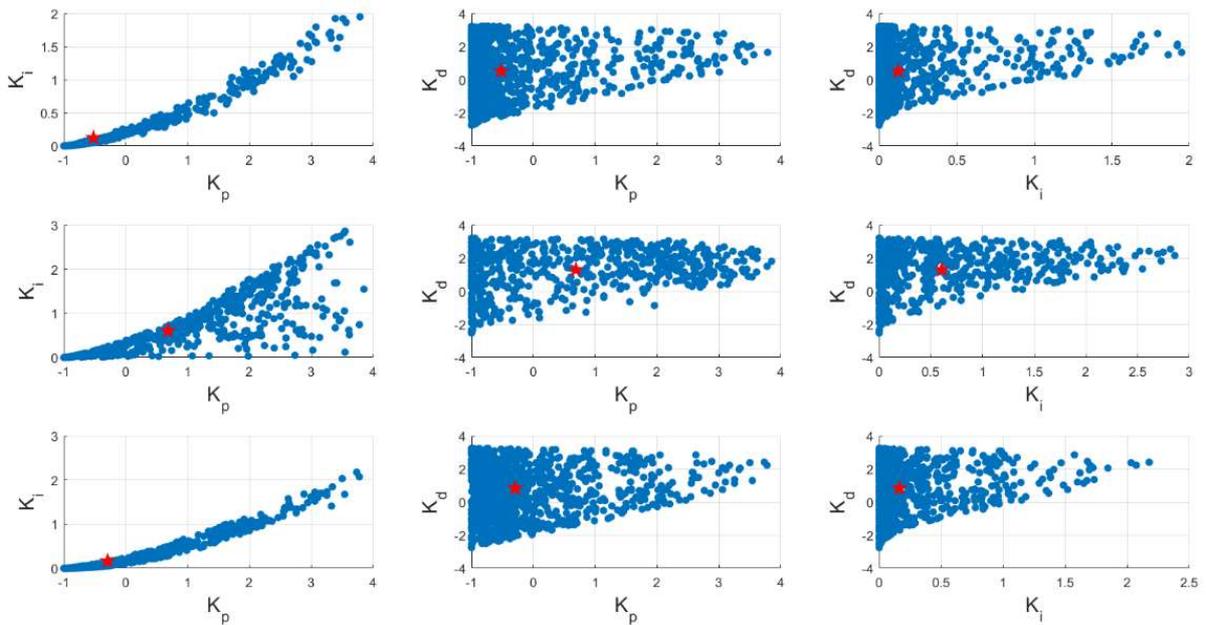

*Figure 6: Stability regions of PID controller gains to control test-bench plant $G_2$ using different non-dominant pole types as (top) all complex poles, (middle) all real poles, (bottom) two real and two complex poles.*

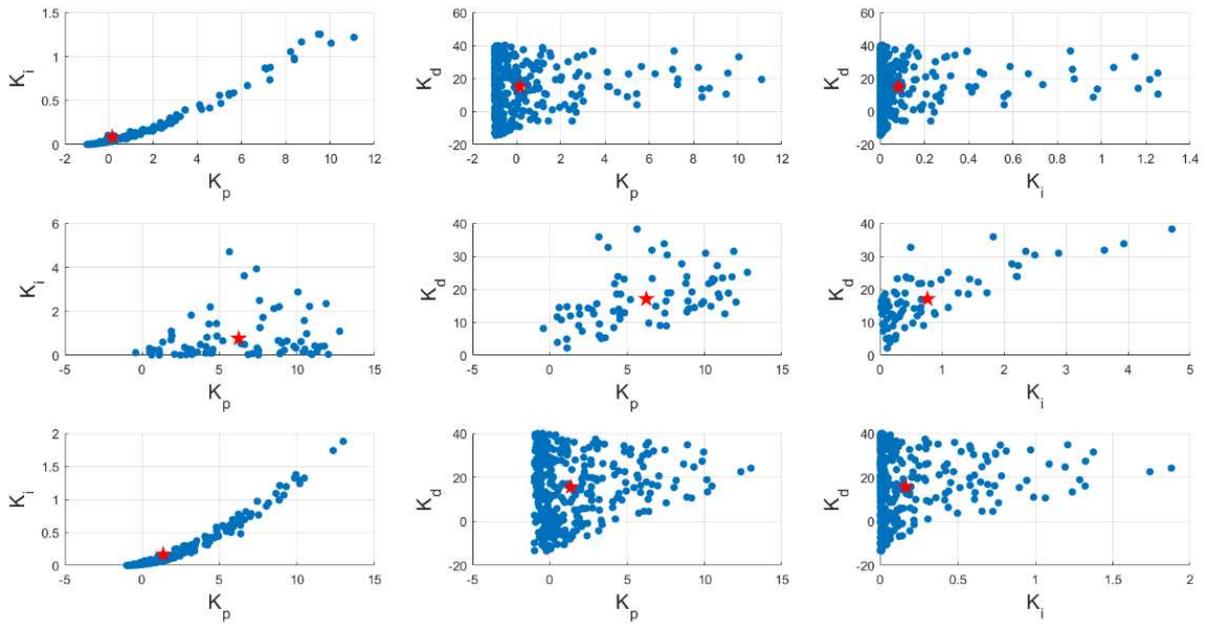

*Figure 7: Stability regions of PID controller gains to control test-bench plant $G_3$ using different non-dominant pole types as (top) all complex poles, (middle) all real poles, (bottom) two real and two complex poles.*

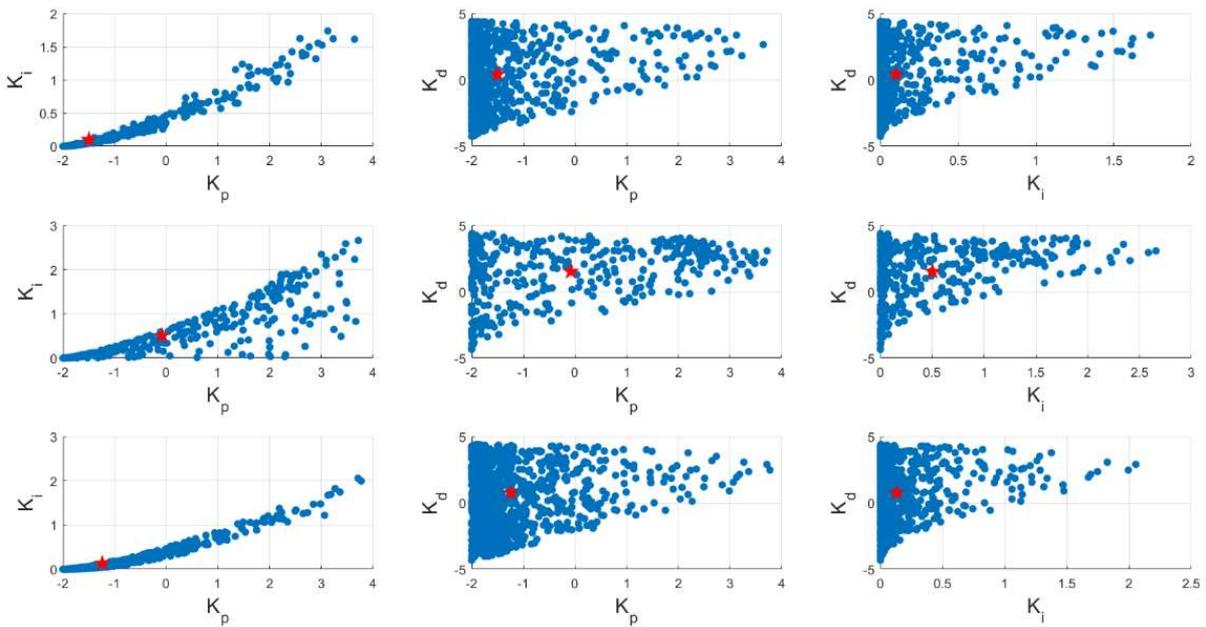

*Figure 8: Stability regions of PID controller gains to control test-bench plant $G_4$ using different non-dominant pole types as (top) all complex poles, (middle) all real poles, (bottom) two real and two complex poles.*

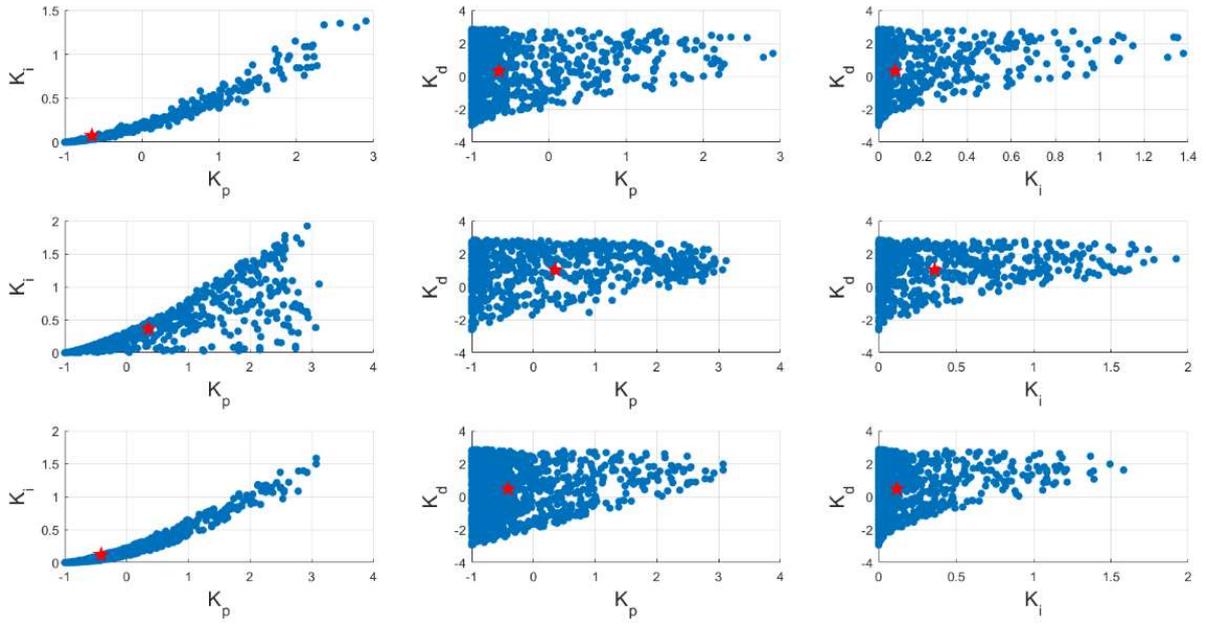

*Figure 9: Stability regions of PID controller gains to control test-bench plant $G_5$ using different non-dominant pole types as (top) all complex poles, (middle) all real poles, (bottom) two real and two complex poles.*

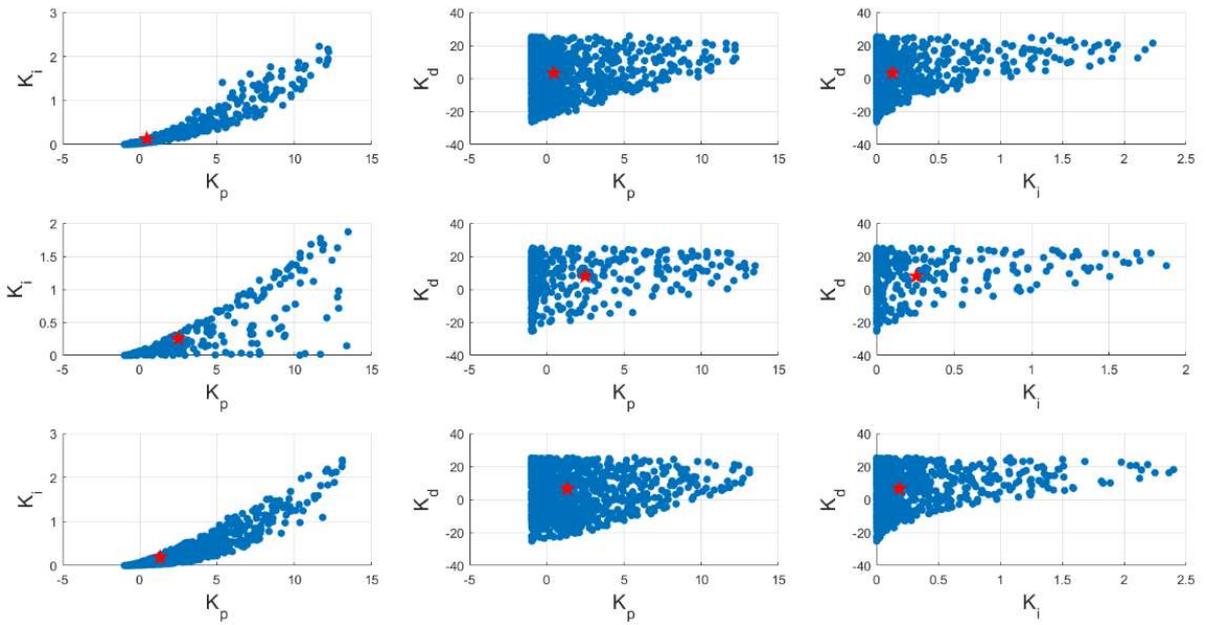

*Figure 10: Stability regions of PID controller gains to control test-bench plant $G_6$ using different non-dominant pole types as (top) all complex poles, (middle) all real poles, (bottom) two real and two complex poles.*

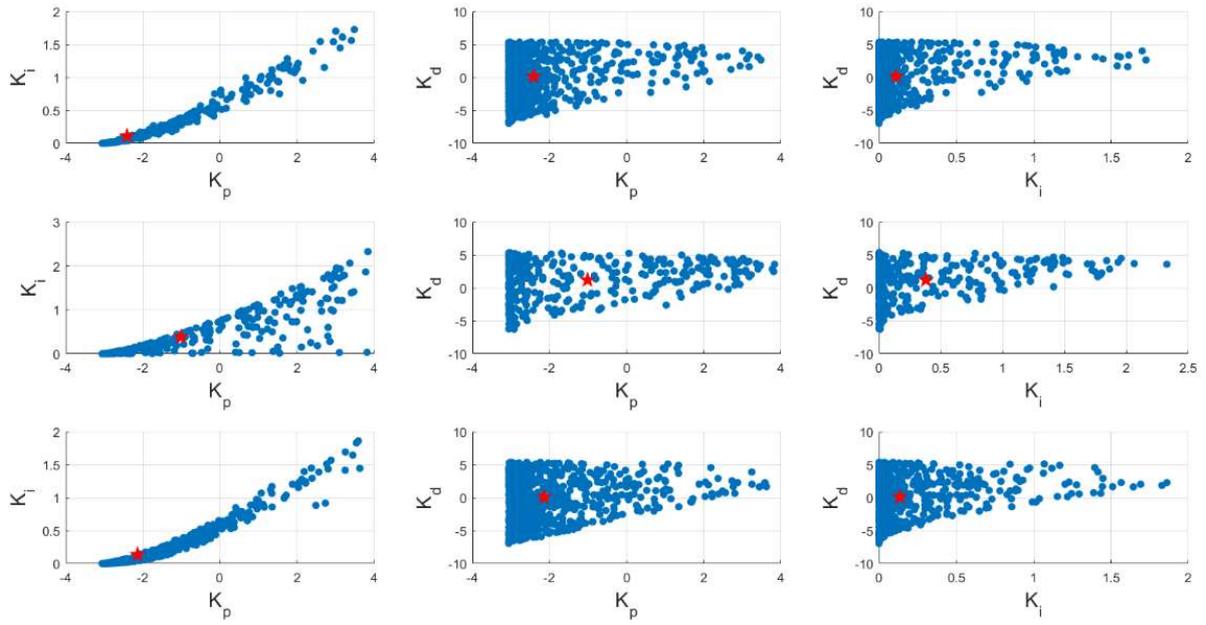

*Figure 11: Stability regions of PID controller gains to control test-bench plant $G_7$ using different non-dominant pole types as (top) all complex poles, (middle) all real poles, (bottom) two real and two complex poles.*

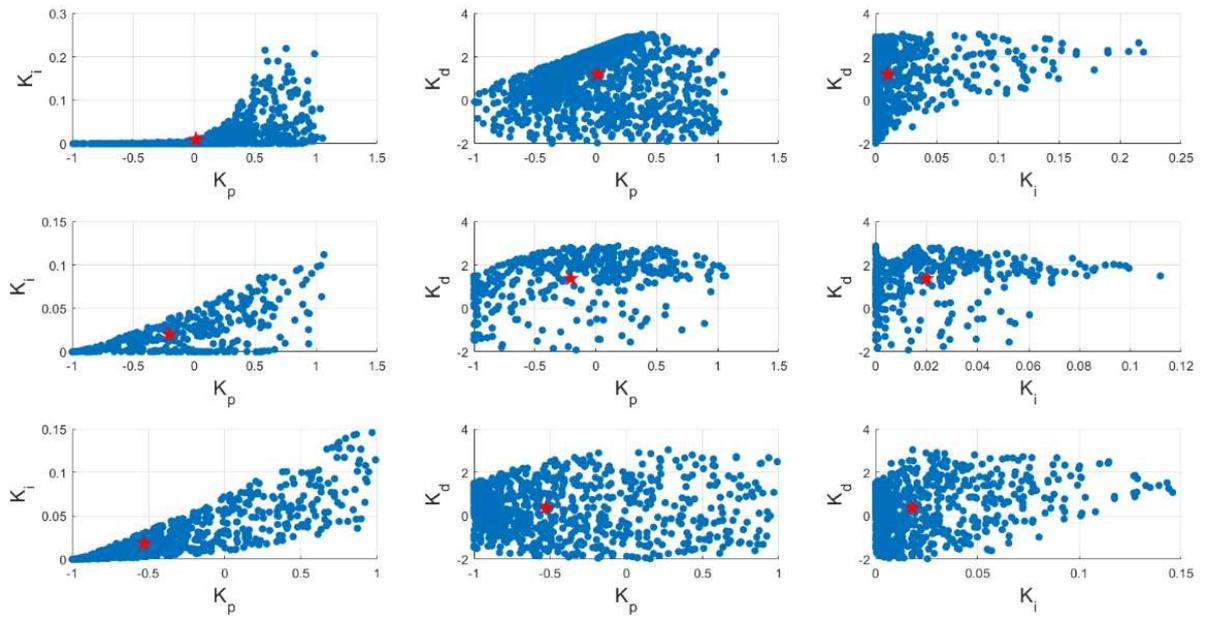

*Figure 12: Stability regions of PID controller gains to control test-bench plant $G_8$ using different non-dominant pole types as (top) all complex poles, (middle) all real poles, (bottom) two real and two complex poles.*

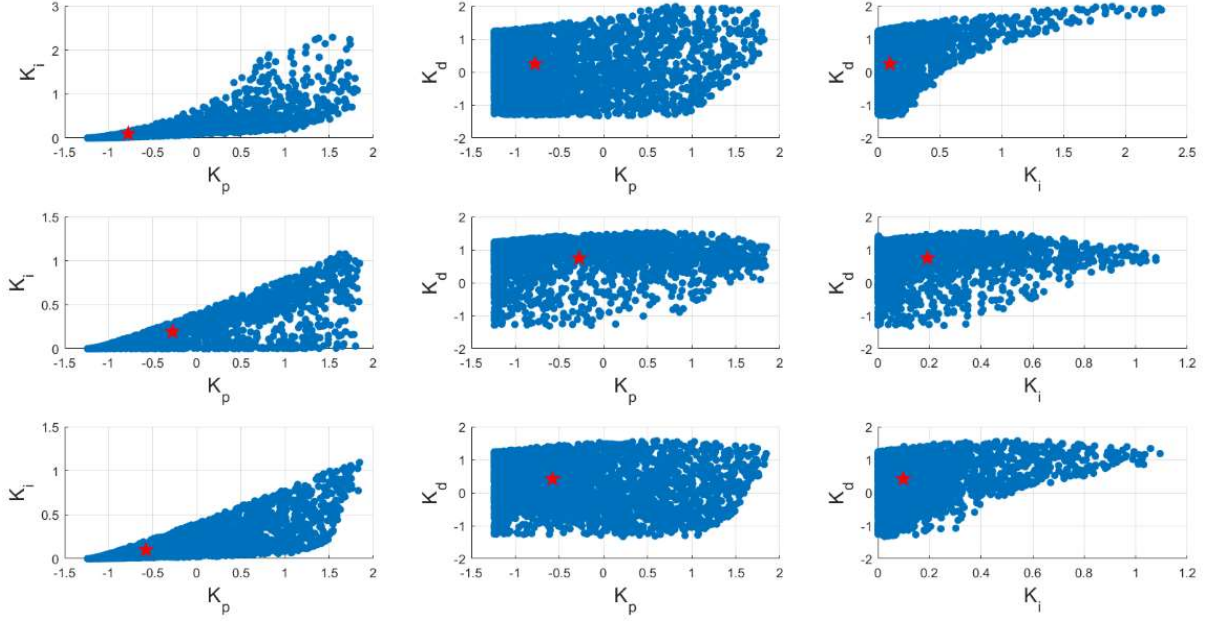

*Figure 13: Stability regions of PID controller gains to control test-bench plant G$_9$ using different non-dominant pole types as (top) all complex poles, (middle) all real poles, (bottom) two real and two complex poles.*

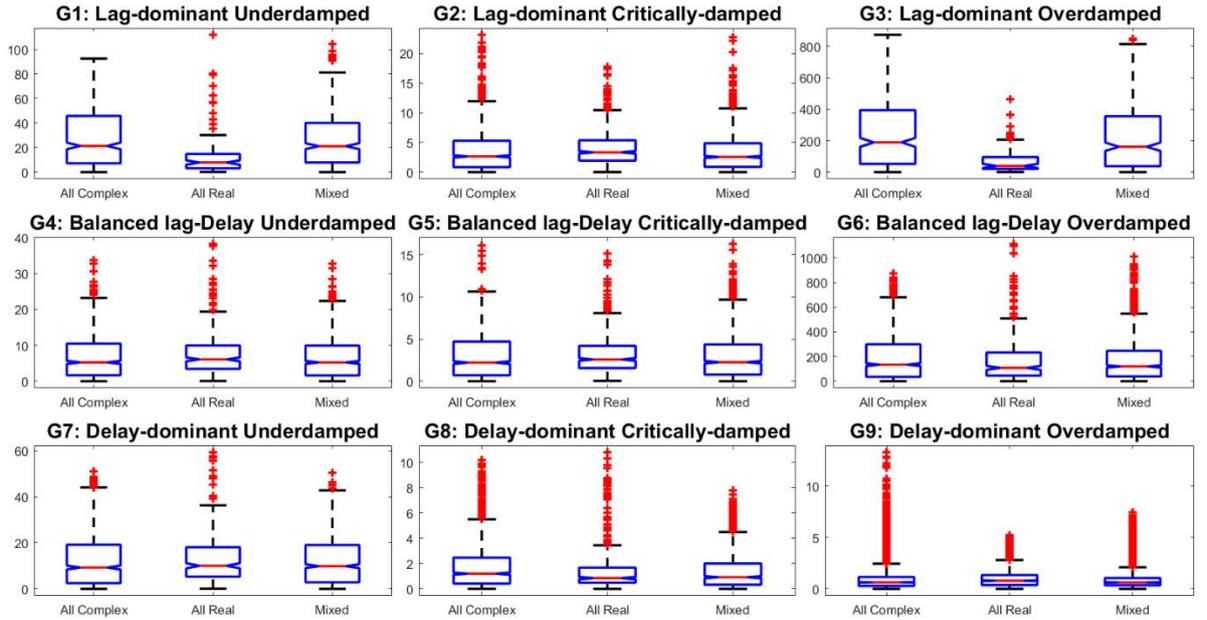

*Figure 14: Distance of the stabilizing samples from the centroid for the nine test-bench processes, indicating compactness of the clusters of data points in the respective stability regions.*

### 4.3. Performance of the robust stable solutions for the test-bench SOPTD processes

In this section, the control performance of the robust stable solutions are quantified for all the nine test-bench processes using the signal norms ($\mathcal{L}_2/\mathcal{L}_\infty$) or system norms ($\mathcal{H}_2/\mathcal{H}_\infty$) of different sensitivity functions, as introduced in section 3.2. The performances are quantified in terms of the disturbance rejection ($d$), control effort ($u$), measurement noise filtering ($n$), tracking error ($e$), gain margin ($G_m$), phase margins ($\Phi_m$) and gain cross-over frequency ($\omega_{gc}$). Many recent literatures argued that in PID control loops, the disturbance rejection and the control signal affecting the actuator size are considered to be the two most significant criteria (Doyle et al. 2013). However apart from the sensitivity function based 2/∞-norms, the traditional performance measures like $\{G_m, \Phi_m, \omega_{gc}\}$ are also compared in Table 3 that may help in understanding the robustness to gain variations, oscillations and the speed of the closed loop system respectively, for different robust stable solutions corresponding to the three non-dominant pole types.

*Table 3: Control performance with the robust stable PID controller gains for the nine classes of test-bench processes with fixed process parameters*

| Process type | Open loop damping | Nature of non-dominant pole | Performance measures of the robust stable PID controller with nominal process parameters | | | | | | | | | | |
|---|---|---|---|---|---|---|---|---|---|---|---|---|---|
| | | | $J_2^d$ | $J_\infty^d$ | $J_2^u$ | $J_\infty^u$ | $J_2^n$ | $J_\infty^n$ | $J_2^e$ | $J_\infty^e$ | $G_m$ | $\Phi_m$ | $\omega_{gc}$ |
| Lag dominant | Underdamped | all complex | 3.3044 | 19.9771 | 82.7783 | 4.4093 | 0.7415 | 1.0458 | 3.2822 | 19.1229 | 2.3691 | 57.1641 | 0.0624 |
| | | all real | 0.4736 | 1.2697 | 65.8704 | 9.4701 | 1.2351 | 2.2471 | 1.3674 | 2.5484 | 1.7022 | 28.6972 | 0.9056 |
| | | two complex/ two real | 1.4191 | 6.0633 | 56.2284 | 6.6734 | 0.8891 | 1.3416 | 1.6370 | 5.7103 | 2.0109 | 54.0378 | 0.8813 |
| | Critically-damped | all complex | 3.3439 | 14.0064 | 78.7187 | 1.4461 | 0.5511 | 2.0063 | 3.5184 | 14.4937 | 1.6132 | 38.4221 | 0.1305 |
| | | all real | 0.8524 | 2.0666 | 37.1731 | 2.3733 | 0.8002 | 1.0695 | 1.1890 | 2.3550 | 2.4179 | 62.6255 | 0.4096 |
| | | two complex/ two real | 2.3366 | 8.3256 | 67.7462 | 1.2429 | 0.5160 | 1.2158 | 2.5147 | 8.6536 | 2.5592 | 49.7392 | 0.1481 |
| | Overdamped | all complex | 2.5287 | 20.5746 | 76.7873 | 5.7889 | 0.4322 | 1.1142 | 3.1520 | 23.3761 | 2.6707 | 56.9813 | 0.0459 |
| | | all real | 0.3742 | 1.3103 | 70.3277 | 13.5142 | 0.7863 | 1.7215 | 1.8449 | 4.1672 | 1.8411 | 37.6897 | 0.4085 |
| | | two complex/ two real | 1.2487 | 7.3045 | 37.4806 | 7.4410 | 0.4795 | 1.0250 | 2.0481 | 8.9502 | 2.4889 | 90.0900 | 0.0897 |
| Balanced lag and delay | Underdamped | all complex | 3.6135 | 17.8317 | 177.2366 | 3.5547 | 0.8981 | 4.1980 | 7.2048 | 35.4755 | 1.2433 | 32.0756 | 0.0740 |
| | | all real | 0.8637 | 2.3685 | 78.2454 | 2.2434 | 0.5553 | 1.0199 | 1.7981 | 4.6522 | 3.0004 | 59.6834 | 0.2182 |
| | | two complex/ two real | 2.5086 | 10.3044 | 137.8486 | 2.7726 | 0.6468 | 2.2735 | 5.0116 | 20.4856 | 1.4595 | 41.4284 | 0.0803 |
| | Critically-damped | all complex | 5.0056 | 24.4705 | 84.7916 | 1.5971 | 0.6311 | 2.8661 | 5.1543 | 24.9874 | 1.3805 | 34.1570 | 0.0926 |
| | | all real | 1.2026 | 3.2422 | 41.9556 | 1.3970 | 0.6216 | 1.0275 | 1.4781 | 3.4699 | 2.7956 | 62.0941 | 0.2782 |
| | | two complex/ two real | 3.0487 | 11.6793 | 63.3211 | 1.2953 | 0.4548 | 1.4710 | 3.2131 | 12.0332 | 1.9285 | 45.5859 | 0.1198 |
| | Overdamped | all complex | 1.9220 | 14.1479 | 94.6679 | 1.8423 | 0.2737 | 1.6764 | 3.7439 | 26.1029 | 7.9973 | 34.8646 | 0.0647 |
| | | all real | 0.8470 | 4.3265 | 57.1623 | 9.4263 | 0.4782 | 1.1939 | 2.2895 | 9.3214 | 3.0638 | 59.7635 | 0.1123 |
| | | two complex/ two real | 1.2354 | 7.4122 | 72.6952 | 6.0139 | 0.3896 | 1.3105 | 2.7746 | 14.7869 | 3.8712 | 49.1662 | 0.0813 |
| Delay dominant | Underdamped | all complex | 3.0527 | 16.3049 | 271.0282 | 5.9197 | 1.0286 | 5.2373 | 9.2686 | 49.4705 | 1.1931 | 29.9875 | 0.0589 |
| | | all real | 0.9421 | 2.9722 | 124.1419 | 3.5254 | 0.4522 | 1.1578 | 2.8602 | 8.9635 | 2.1639 | 53.6541 | 0.1262 |
| | | two complex/ two real | 2.2789 | 10.0616 | 108.0883 | 4.6069 | 0.7963 | 3.2323 | 6.9138 | 30.4903 | 1.3145 | 37.2722 | 0.0618 |
| | Critically-damped | all complex | 7.4984 | 103.7771 | 64.5575 | 1.4312 | 0.7079 | 1.0000 | 7.5587 | 103.7771 | 2.5635 | 84.2525 | 0.0095 |
| | | all real | 6.2557 | 50.2831 | 24.6987 | 1.6705 | 0.8474 | 1.0021 | 6.3342 | 50.2831 | 2.0175 | 64.5762 | 0.0198 |
| | | two complex/ two real | 8.9267 | 71.3828 | 59.5067 | 1.3151 | 0.3884 | 1.8875 | 8.9794 | 71.4827 | 1.5833 | 43.7891 | 0.0209 |
| | Overdamped | all complex | 4.0000 | 19.1321 | 116.2718 | 1.9777 | 0.8174 | 2.8856 | 5.0481 | 23.9850 | 1.3746 | 34.3820 | 0.0965 |
| | | all real | 1.9469 | 6.5425 | 59.9512 | 4.2854 | 1.5701 | 1.1506 | 2.5436 | 8.2736 | 1.9999 | 51.8288 | 0.1446 |
| | | two complex/ two real | 3.1262 | 12.5264 | 77.9316 | 1.5506 | 0.9072 | 1.5167 | 3.9621 | 15.6585 | 1.7872 | 47.1110 | 0.0865 |

The corresponding time domain responses for control variable due to step change in set-point (left top), step change in disturbance input (right top), manipulated variable or the control signal for step change in set-point (left bottom) and control signal for step change in disturbance input (right bottom) are shown in Figure 15-Figure 23, for the nine test-bench processes. Both $J_2^d$ and $J_\infty^d$ are found to be the smallest for the robust stable PID controller with the all real non-dominant pole type, as evident from Table 3 as well as the right top panel of Figure 15-Figure 23.

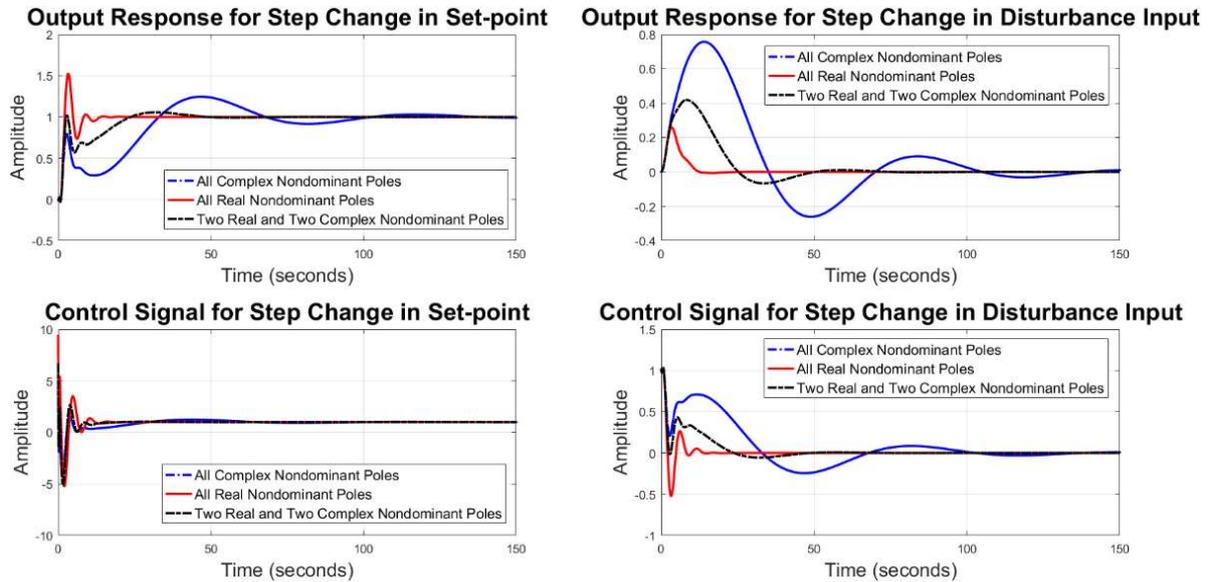

*Figure 15: Controlled variable (top) and manipulated variable (bottom) due to step change in set-point (left) and disturbance input (right) for process $G_1$*

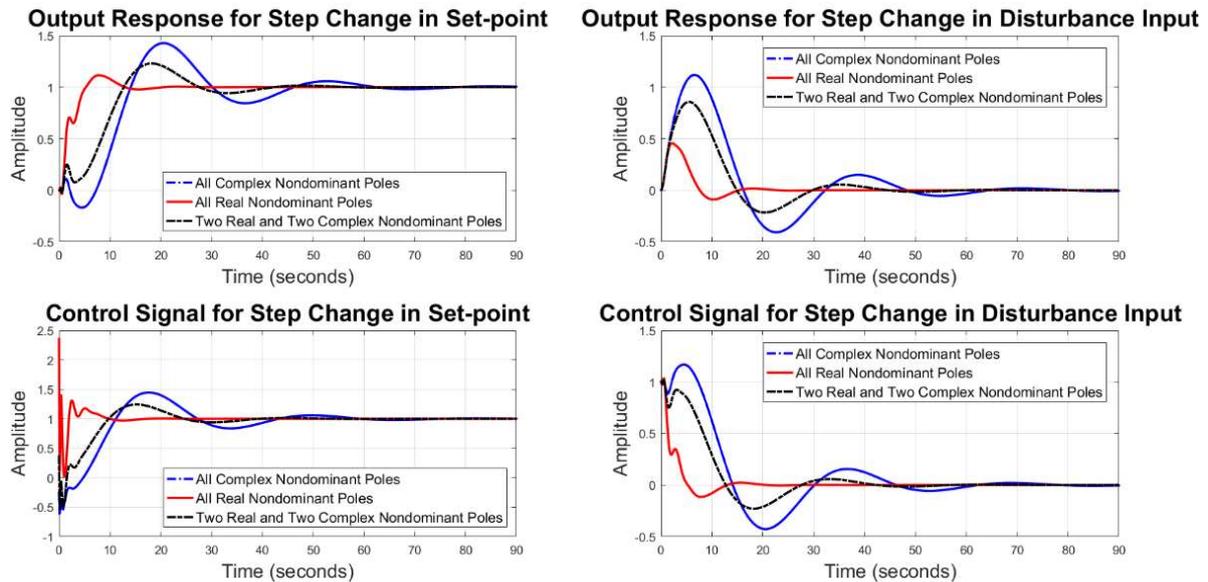

*Figure 16: Controlled variable (top) and manipulated variable (bottom) due to step change in set-point (left) and disturbance input (right) for process $G_2$*

The all complex non-dominant pole case has the worst disturbance rejection performance and the two-complex/two-real case falls in between these two cases. The tracking performance is also the best for the all real non-dominant poles as revealed from the values of $J_2^e$ and $J_\infty^e$ in Table 3 and the responses of the control variables, subjected to a step change in the set-point in the left top panels of Figure 15-Figure 23. For the all complex non-dominant poles, the step response is sluggish and highly oscillatory, amongst the three types of non-dominant poles. However the best tracking and disturbance rejection performance of the all real non-dominant poles comes at the cost of increased actuator size ($J_\infty^u$) or higher control effort ($J_2^u$) or both as reported in Table 3 and the two bottom panels of Figure 15-Figure 23, for different test-bench processes. The noise rejection performance is also not always the best using the all real non-dominant poles, as evident from $J_2^n$ and $J_\infty^n$ in Table 3. The $\omega_{gc}$ in Table 3 is always found to be the highest using the all real non-dominant poles, yielding a faster time response. But this increased speed of the robust stable solution comes at the cost of reduced gain and phase margin ($G_m, \Phi_m$) in some cases. Therefore, as a summary, control systems where the disturbance rejection, speed of set-point tracking are of utmost importance, the all real non-dominant pole based robust stable PID controller can be employed. In addition, the all real non-dominant poles also produce the best phase margin particularly for the balanced lag-delay family of processes, as evident from Table 3.

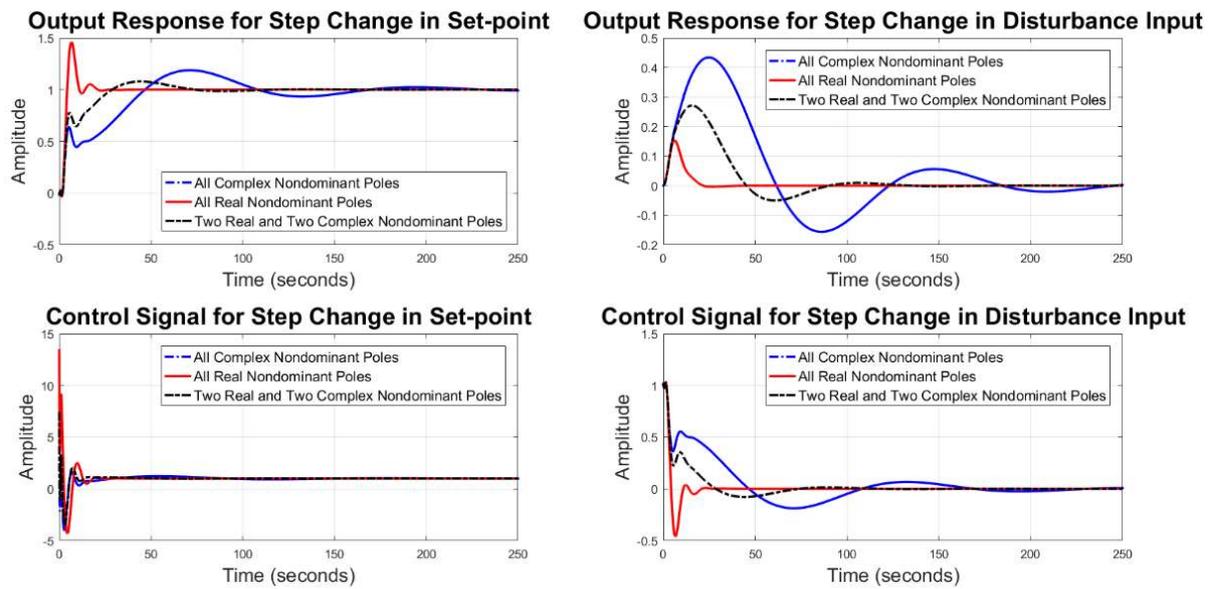

*Figure 17: Controlled variable (top) and manipulated variable (bottom) due to step change in set-point (left) and disturbance input (right) for process $G_3$*

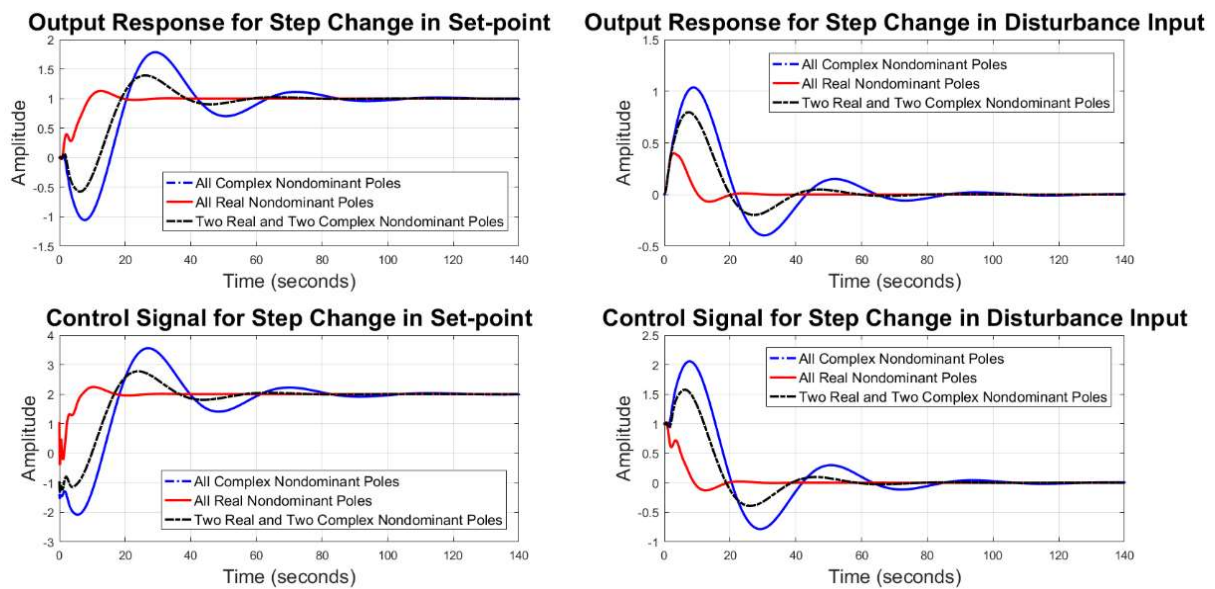

*Figure 18: Controlled variable (top) and manipulated variable (bottom) due to step change in set-point (left) and disturbance input (right) for process $G_4$*

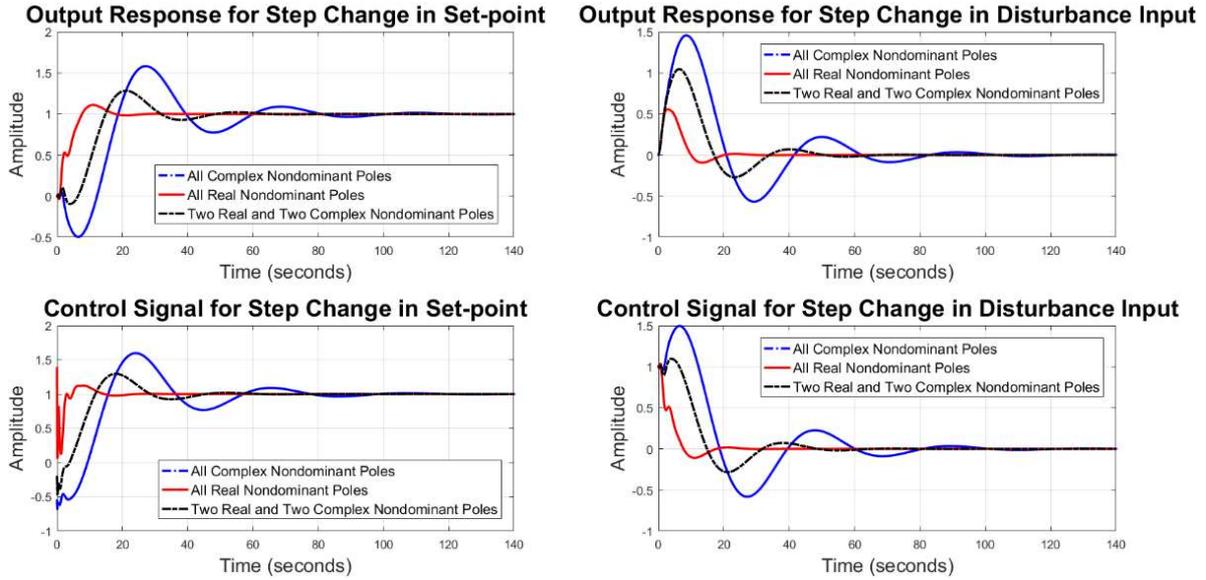

*Figure 19: Controlled variable (top) and manipulated variable (bottom) due to step change in set-point (left) and disturbance input (right) for process $G_5$*

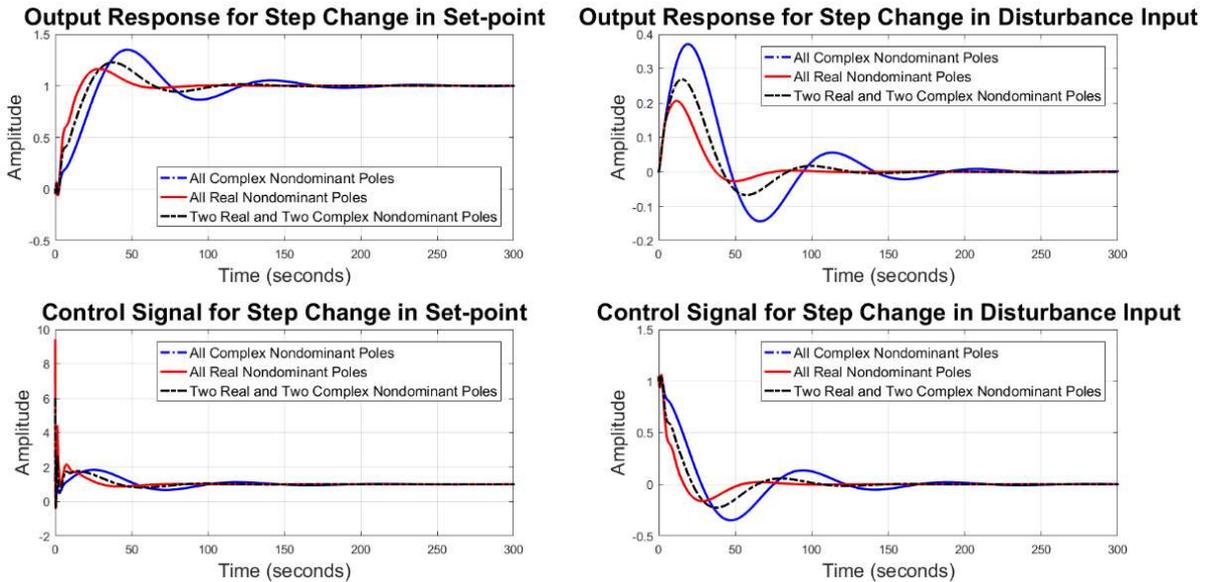

*Figure 20: Controlled variable (top) and manipulated variable (bottom) due to step change in set-point (left) and disturbance input (right) for process $G_6$*

However some of the above mentioned measures in Table 3, may indicate a similar performance and hence a performance correlation analysis as shown in Figure 24 may reveal which measures are inter-related and which are not. A threshold based analysis of the correlation coefficients ($R$) of the robust stable performance measures shows that two pairs *viz.* $\{J_2^d, J_\infty^d\}$ and $\{J_\infty^d, J_\infty^e\}$ are highly correlated with $R>0.9$, hence any one of them within the pair would be a sufficiently independent measure of the closed loop performance. This is inherently different from the trade-off based design approach for PID control loops, where it is usually considered that the performance measures are always independent of each other. Therefore, the correlation plots in Figure 24 for the all real nondominant pole case with best achievable overall performance indicate that improvement in one performance criterion may not always lead to deterioration of the others. In recent years, there have been many works e.g. (Das & Pan 2014)(Herreros et al. 2002; Pan & Das 2012; Pan & Das 2013)(Hajiloo et al. 2012) which have used multi-objective optimization for controller design by minimizing multiple conflicting objectives together. However, in most of these cases, the performance correlation analysis was not shown to illustrate whether a chosen set of cost functions or performance measures are at all independent which is an important criterion to judge before designing such optimization-based control systems.

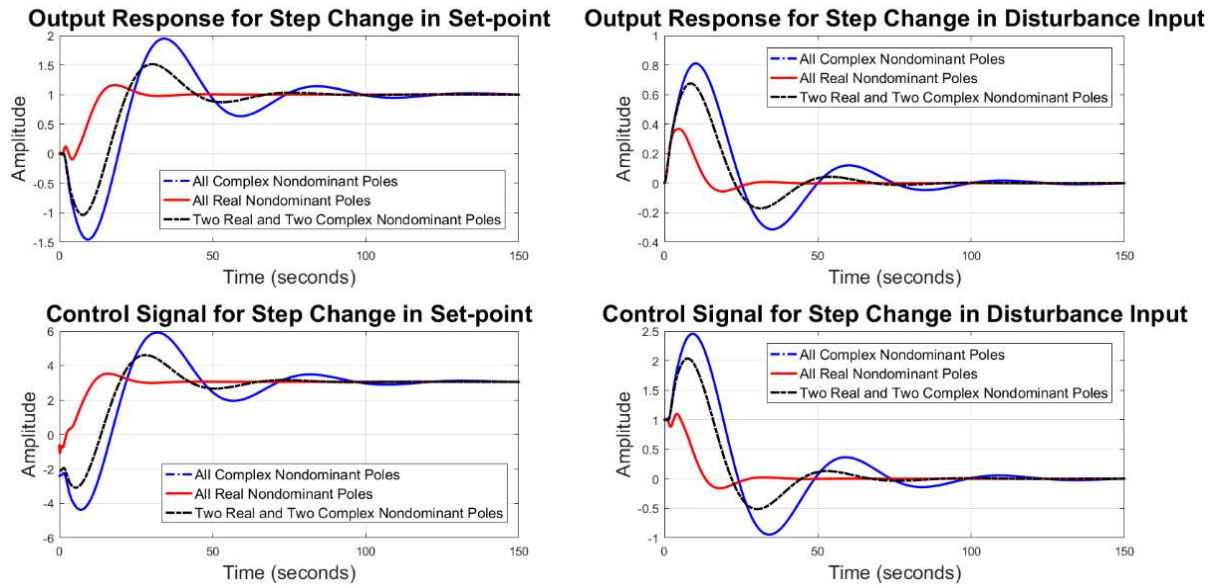

*Figure 21: Controlled variable (top) and manipulated variable (bottom) due to step change in set-point (left) and disturbance input (right) for process $G_7$*

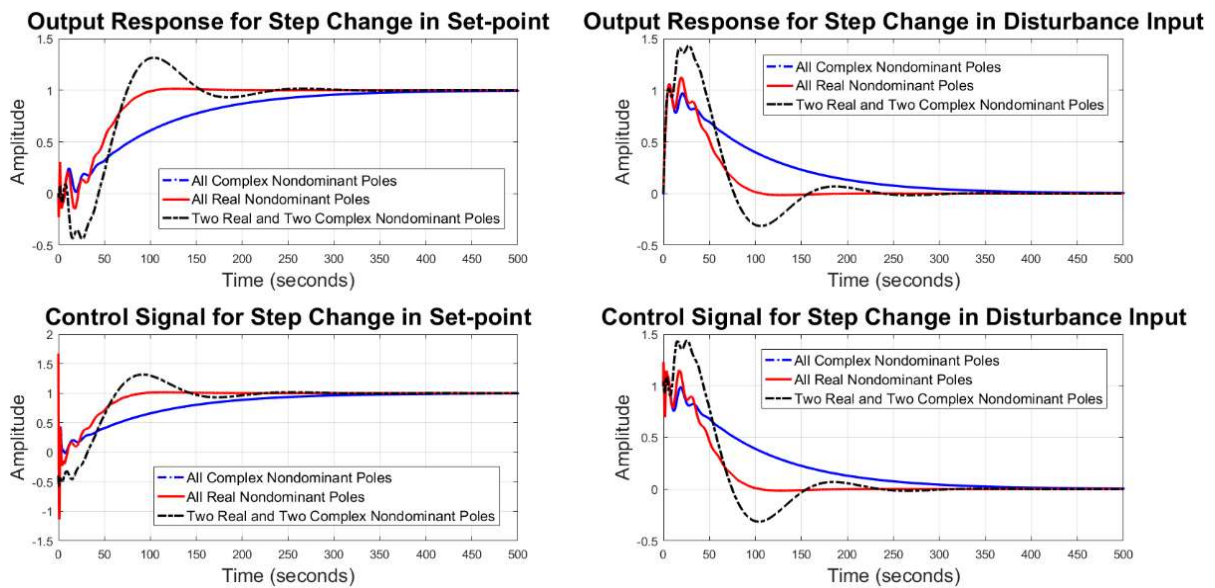

*Figure 22: Controlled variable (top) and manipulated variable (bottom) due to step change in set-point (left) and disturbance input (right) for process $G_8$*

The performance correlation analysis in Figure 24 is particularly important in controller design tasks since there are many recent attempts with an aim of optimization based PID controller tuning using clearly redundant and highly correlated performance measures or cost functions which should have otherwise gone through such performance correlation analysis first, before applying heuristic optimization algorithms on them and comparing marginal performance improvement amongst various competing global optimizers. As prominent examples of such redundant and correlated cost function based PID controller design can be noted in (Panda et al. 2012) and elsewhere. In particular, some approaches reported a comparison of different cost functions with different units and also having different ranges which are unjustified e.g. in (Panda et al. 2012). Moreover, for time domain tracking comparisons, usually integral performance indices like integral of absolute error (IAE), integral of squared error (ISE), integral of time multiplied absolute error (ITAE) and integral of time multiplied squared error (ITSE) capture the combined effects of overshoot, steady state error, rise-time, settling-time and peak-time and do not need to be added separately in the cost function unlike the approaches reported in many recent works e.g. (dos Santos Coelho 2009)(Zhu et al. 2009)(Ramezanian et al. 2013)(Sahib 2015)(Bendjeghaba 2014)(Zeng et al. 2015). Therefore including these time domain features either in the cost function along with an integral performance criteria can be considered redundant and unnecessary as previously adopted in (Zamani et al. 2009)(Aguila-Camacho & Duarte-Mermoud 2013) or formulating customized cost functions as weighted average of many correlated cost functions (Das et al. 2011).

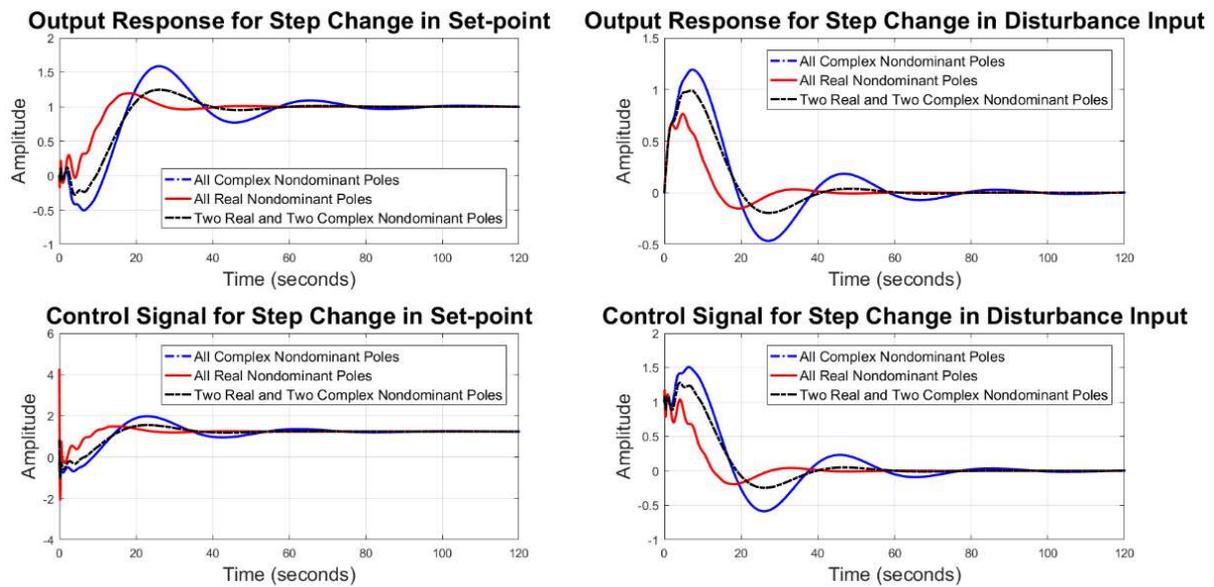

*Figure 23: Controlled variable (top) and manipulated variable (bottom) due to step change in set-point (left) and disturbance input (right) for process $G_9$*

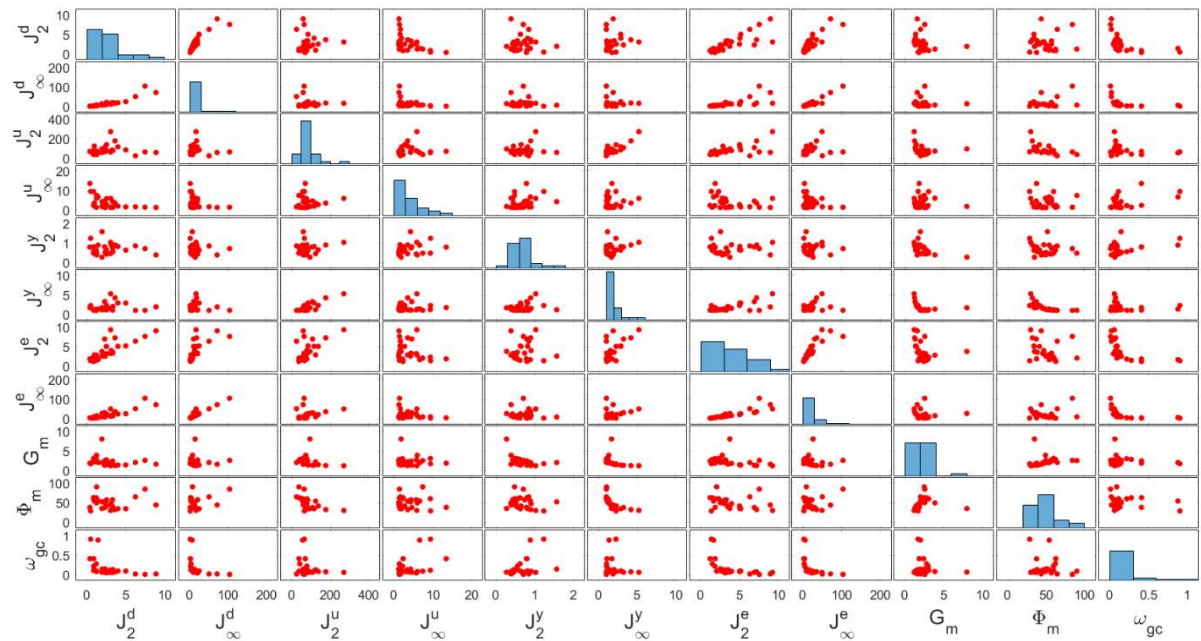

*Figure 24: Correlation analysis of the 11 performance measures for the robust stable solutions of the test-bench processes.*

In some cases, performance criteria in time *vs.* frequency domain can be expected to be highly correlated to each other from physical understanding of the underlying control principles e.g. overshoot *vs.* phase margin (Zamani et al. 2009). Another optimization based approaches in (Hasanien 2013) also reported tuning subset of controller parameters leaving the integral gain, optimization on fitted polynomials sampled on limited number of points in the search space, and even reporting non-zero steady state error in PID control loops which are unrealistic and limiting. However, our approach is not only an improvement over these literatures, reporting different system/signal norms for the robust stable solutions using a standardised set of traditional control loop performance measures but also investigates how much redundant information is embedded within these chosen set of control performance quantification measures. Various performance measures were also previously used in (Zamani et al. 2017)(Chen et al. 2014)(Das & Pan 2014)(Pan & Das 2012; Pan & Das 2013), where explanations for such possible conflicts amongst a chosen set of cost functions can be found from physical understanding of the controlled, manipulated and other state variables as well as their role in the cost function formulation e.g. speed (gain crossover frequency) *vs.* overshoot (phase margin) trade-off, tracking *vs.* control effort trade-off, servo (tracking) *vs.* regulatory (disturbance rejection) trade-off etc. (Das & Pan 2014)(Alcántara et al. 2013). However, for multiple objective based control system design, a negatively correlated pattern or conflicting nature amongst the chosen cost functions is a necessary criterion to balance various hidden aspects of the overall achievable control performance, apart from finding only robust stable solutions what the present paper proposes.

### *4.4. Invariance of the order of Pade approximation on the control performance*

Next, we show the effect of changing the order of Pade approximation using the best found robust controller i.e. using the all real non-dominant pole configuration which is found to have the fastest set-point tracking as well as disturbance rejection performances. The Pade approximation for the time delay term $e^{-Ls}$ can be represented in a general form (Silva et al. 2007), with user defined order ($N_{Pade}$) as:

$$\begin{aligned} e^{-Ls} &\cong N_r(Ls)/D_r(Ls), \\ N_r(Ls) &= \sum_{k=0}^{r} \frac{(2r-k)!}{k!(r-k)!}(-Ls)^k, \\ D_r(Ls) &= \sum_{k=0}^{r} \frac{(2r-k)!}{k!(r-k)!}(Ls)^k, \end{aligned} \quad (34)$$

where $r$ is the order of the Pade approximation i.e. $N_{Pade} = r$.

Using (34), the Pade approximation for different orders of $N_{Pade} = \{3, 5, 7, 9\}$ becomes:

$$e^{-Ls}_{N_{Pade}=3} \simeq \frac{-L^3s^3 + 12L^2s^2 - 60L + 120}{L^3s^3 + 12L^2s^2 + 60L + 120}$$

$$e^{-Ls}_{N_{Pade}=5} \simeq \frac{-L^5s^5 + 30L^4s^4 - 420L^3s^3 + 3360L^2s^2 - 15120Ls + 30240}{L^5s^5 + 30L^4s^4 + 420L^3s^3 + 3360L^2s^2 + 15120Ls + 30240}$$

$$e^{-Ls}_{N_{Pade}=7} \simeq \frac{-L^7s^7 + 56L^6s^6 - 1512L^5s^5 + 25200L^4s^4 - 277200L^3s^3 + 1995840L^2s^2 - 8648640Ls + 17297280}{L^7s^7 + 56L^6s^6 + 1512L^5s^5 + 25200L^4s^4 + 277200L^3s^3 + 1995840L^2s^2 + 8648640Ls + 17297280}$$

$$e^{-Ls}_{N_{Pade}=9} \simeq \frac{\begin{pmatrix} -L^9s^9 + 90L^8s^8 - 3960L^7s^7 + 110880L^6s^6 - 2162160L^5s^5 + 30270240L^4s^4 - 302702400L^3s^3 \\ +2075673600L^2s^2 - 8821612800Ls + 1.76432256\times 10^{10} \end{pmatrix}}{\begin{pmatrix} L^9s^9 + 90L^8s^8 + 3960L^7s^7 + 110880L^6s^6 + 2162160L^5s^5 + 30270240L^4s^4 + 302702400L^3s^3 \\ +2075673600L^2s^2 + 8821612800Ls + 1.76432256\times 10^{10} \end{pmatrix}}.$$

(35)

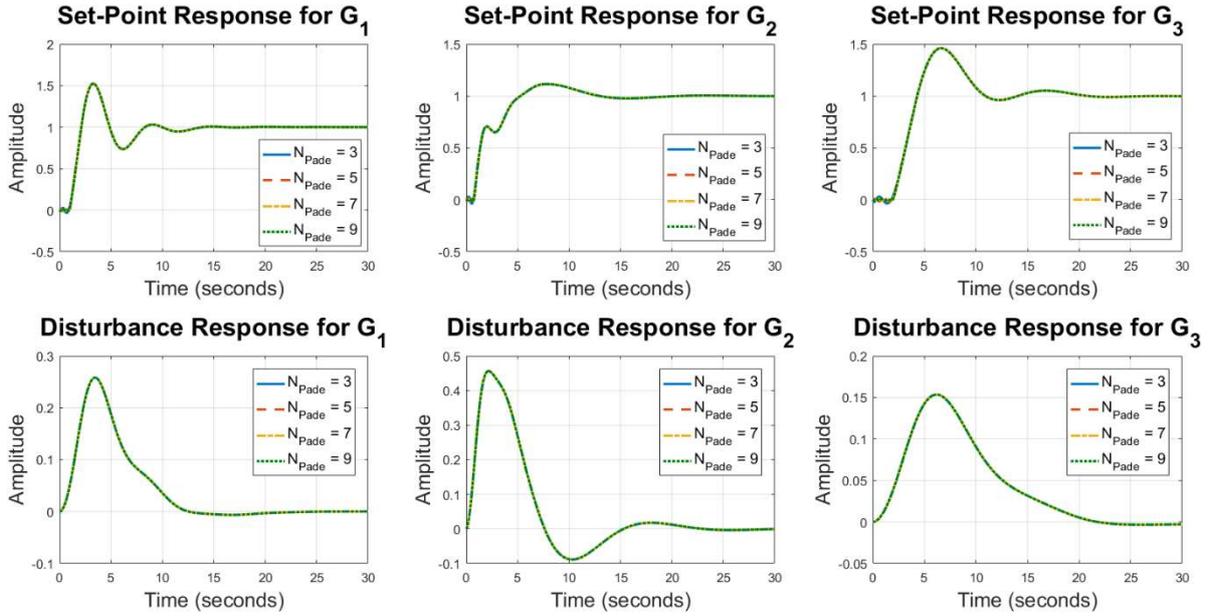

*Figure 25: Set-point and disturbance response for the lag-dominant processes with different order of Pade approximation.*

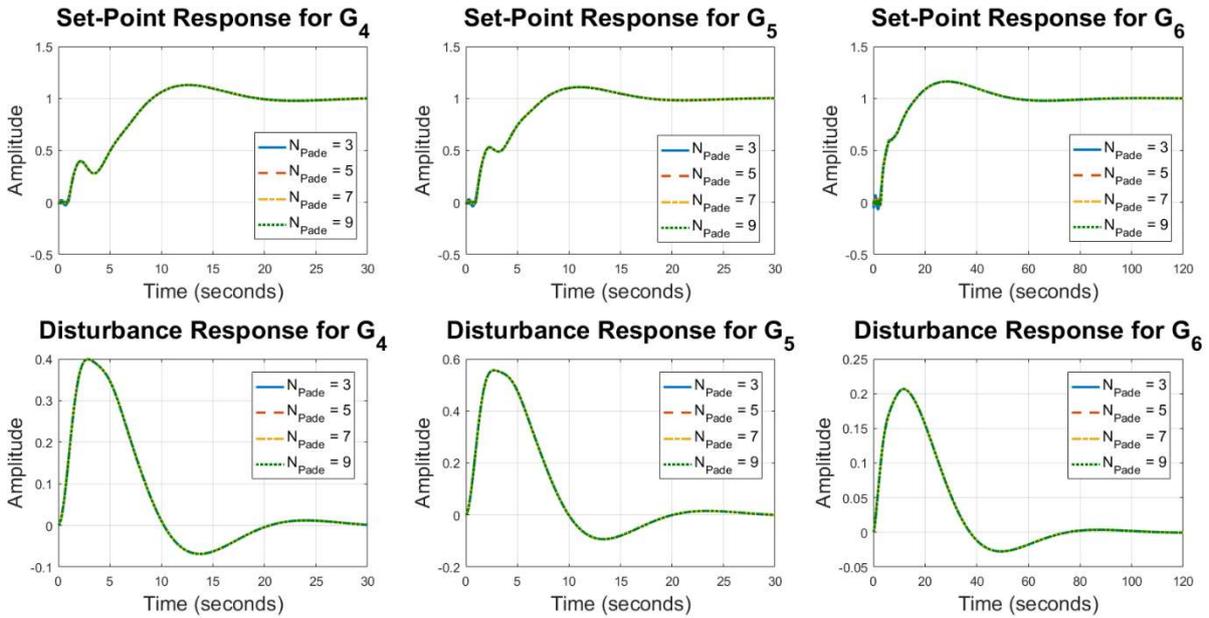

*Figure 26: Set-point and disturbance response for the balanced lag-delay processes with different order of Pade approximation.*

Now, the time delay term of all the nine types of test-bench processes have been approximated by (35). Approximation of the delay term using (35) not only increases the overall system order but also changes the process dynamics due to increased number of root locus branches. Now by increasing the Pade order as $N_{Pade} = \{3, 5, 7, 9\}$, the nine test-bench processes are controlled by the respective robust PID controllers which was originally designed using a third order approximation ($N_{Pade} = 3$) and the deviation in the performance measures have also been noted. The set-point tracking and disturbance rejection tasks for step input remain almost unchanged with the robust-stable PID controller even with higher order of Pade approximation which are shown for the three classes of processes in Figure 25-Figure 27 respectively. The lag-dominant and balanced lag-delay plants show almost imperceptible difference with increasing order of $N_{Pade}$. However, for delay dominant processes, at the initial dead time phase of a step-input response, the number of peaks and troughs in the initial oscillations might vary depending on the delay approximation order $N_{Pade}$, although the overall dynamical characteristics do not change significantly due to the dominant pole placement criteria imposed in the controller design phase.

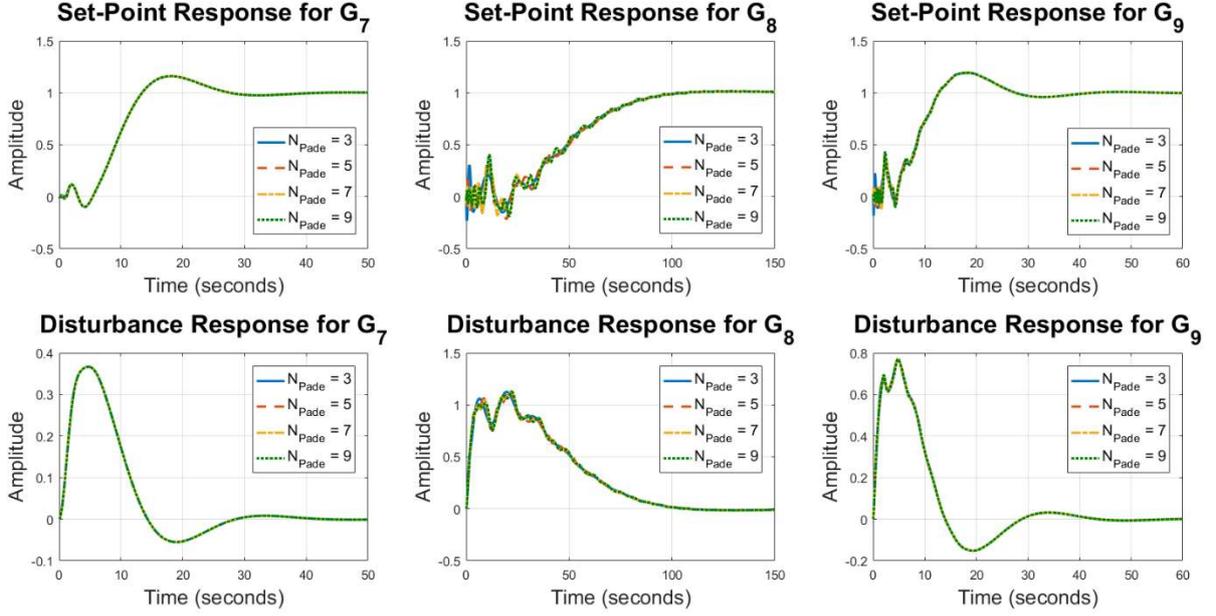

*Figure 27: Set-point and disturbance response for the delay-dominant processes with different order of Pade approximation.*

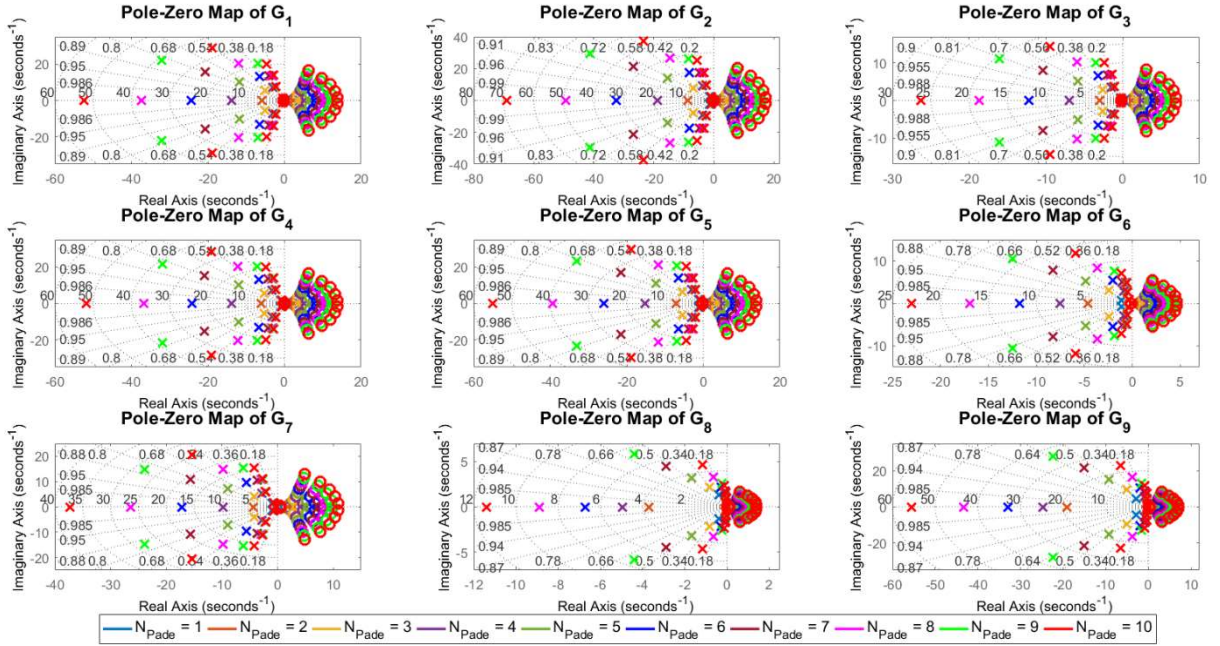

*Figure 28: Closed loop pole-zero maps for dominant pole placement of test-bench processes with increasing order of Pade approximation $N_{Pade}$ = 1 to 10.*

Although in the design process the four roots of equation (10) are considered to be real, however due to the over-determined nature of the problem having 4 alternative solutions for $K_p$ in (12), this would yield different stability regions, amongst which the largest one is selected as discussed before. These sets of stabilizing set of design parameters $\{m, \zeta_{cl}, \omega_{cl}\}$ are then mapped on to the controller parameter $\{K_p, K_i, K_d\}$ space where the clustering is carried out to find out the robust stable solution with the largest stability region. However, using this controller gain does not always guarantee that the characteristic polynomial involving open loop process parameters and three controller gains upon factorization, would always yield real poles. The assumption of different type of non-dominant poles are just to get many alternative candidate expressions for controller gains, so that one can choose the most robust stable solution while also maintaining acceptable closed loop performance.

One of the biggest hurdles of pole placement controller design for time delay systems is any parameterized approach of coefficient matching of the characteristic polynomial is usually infeasible, due to having infinite number of roots. Therefore, the approximation method and order for the delay term need to adapt to the unknown number of closed loop poles. In order to keep the dominant dynamics intact upon variable order of Pade approximation, the effect of the rest of closed loop poles can be minimized using the non-dominance criteria, described above. To verify this, we have shown here the closed loop pole locations with different orders of Pade approximation. The true delay manifests itself as a very high order transfer function, thus creating many interlaced poles and zeros which the PID controller design should be able to handle as reported in Figure 28 for $N_{Pade}$ varying from 1 to 10. It is clear from the pole-zero map in Figure 28 that the non-dominant pole is far away from the rest as expected and the closed loop poles with increasing order of Pade approximation move almost along a constant damping line such that the closed loop performances remain unaltered which are also revealed from the set-point tracking and disturbance rejection responses in time domain (Figure 25-Figure 27). It is also observed from Figure 28 that with the even order of $N_{Pade}$, the non-dominant pole becomes real and for odd $N_{Pade}$, the non-dominant poles become complex conjugates, creating different patterns of closed loop pole locations for even and odd orders of $N_{Pade}$.

### *4.5. Performance of the robust stable controller with perturbed plant parameters*

In this section we now investigate the performance variations considering the process parameters to be uncertain around the robust stable solutions which were found as the centroids of the stability regions using *k*-means clustering. This was obtained by random sampling of the design parameter space $\{m, \zeta_{cl}, \omega_{cl}\}$ and then mapping the accepted stable solutions in the three PID controller parameter space $\{K_p, K_i, K_d\}$, followed by applying *k*-means clustering algorithm to find out the centroid of the stability region in the 3D space of controller gains. These robust stable solutions are considered as the mid-points of the images showing the iso-performance contours where the mid-point represent the performance with fixed robust stable PID controller gains on the nominal process model parameters as reported in section 4.1. Moving towards a better performance as often done in optimization based controller design approaches using single or multiple control objectives may not necessarily yield a robust stable solution (Pan & Das 2016). Rather here we take a different route and design a robust stable solution first (as the centroid of the stability region in controller parameter space) followed by investigating the resulting performance improvement or degradation for uncertain process parameters using this robust stable solution. The maximum allowable positive/negative perturbation in all the parameters are determined until at least one out of the 1000 randomly sampled values yields unbounded performance measure i.e. 2/∞-norms or gain/phase margins, within a chosen range of process parameters around the robust stable point (designed with the fixed nominal process model parameters) i.e. $L \in \left[L^{\min}, L^{\max}\right], T \in \left[T^{\min}, T^{\max}\right], \zeta_{ol} \in \left[\zeta_{ol}^{\min}, \zeta_{ol}^{\max}\right]$.

This is justified from the fact that for linear control systems, the parametric uncertainty of the process under control can be equivalently expressed as uncertainty in the controller gains (Silva et al. 2007). Therefore, a robust stable solution at the centroid of the controller parameter space is expected to handle equivalent model parameter uncertainties, around its nominal values. It has been found that for the lag-dominant processes a maximum of 40-60% perturbations in all three parameters $\{L, T, \zeta_{ol}\}$ can be allowed without encountering an unstable solution out of 1000 randomly sampled points. Also, for balanced lag-delay processes a range of maximum 50-70% perturbation can be allowed. However, this range significantly reduces to 20-30% for delay dominant processes, since stabilizing these processes are more difficult. The variation with lag ($T$) or open loop natural frequency ($\omega_{ol}$) with other parameters like $\{L, \zeta_{ol}\}$ have been found to be smoother compared to the other combinations. The highly heterogeneous performance contours are found for uncertain parameter pairs of $\{L, \zeta_{ol}\}$ where using the robust stable PID controller, the performance contours has many local maxima or minima as revealed from Figure 29-Figure 37 for the nine test-bench processes. In all the contour plots a reddish hue represent a higher value and bluish hue represent lower value.

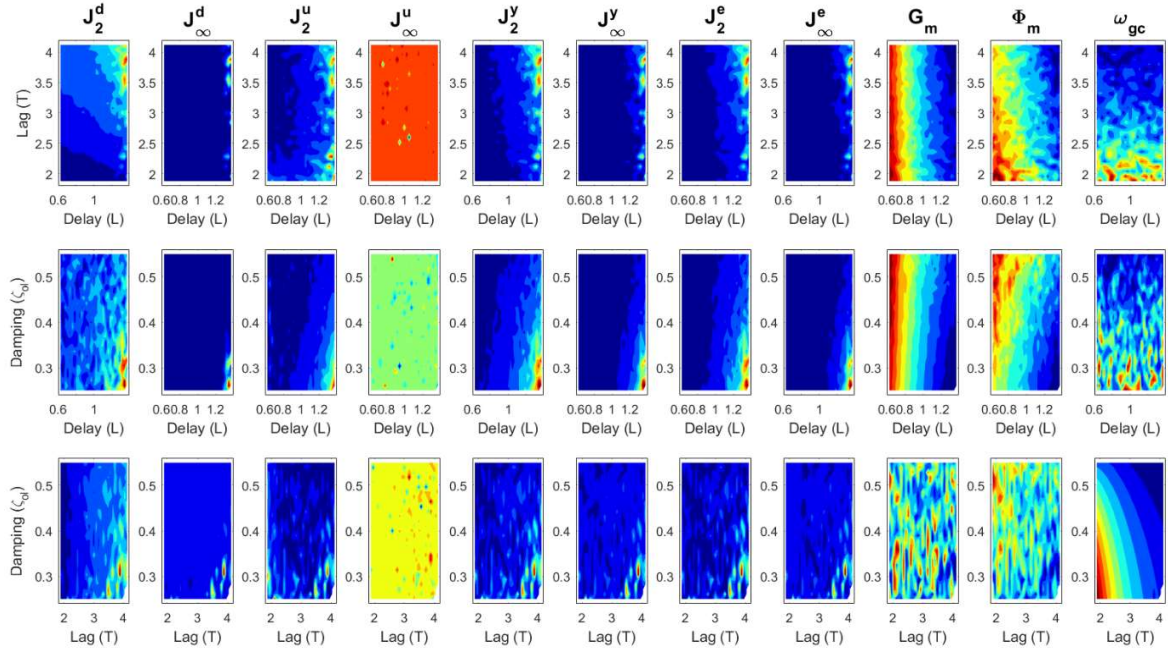

Figure 29: Iso-performance contours as joint distribution of 40% perturbed process parameters {L, T, $\zeta_{ol}$} with the robust stable PID controller having all real non-dominant poles for process $G_1$.

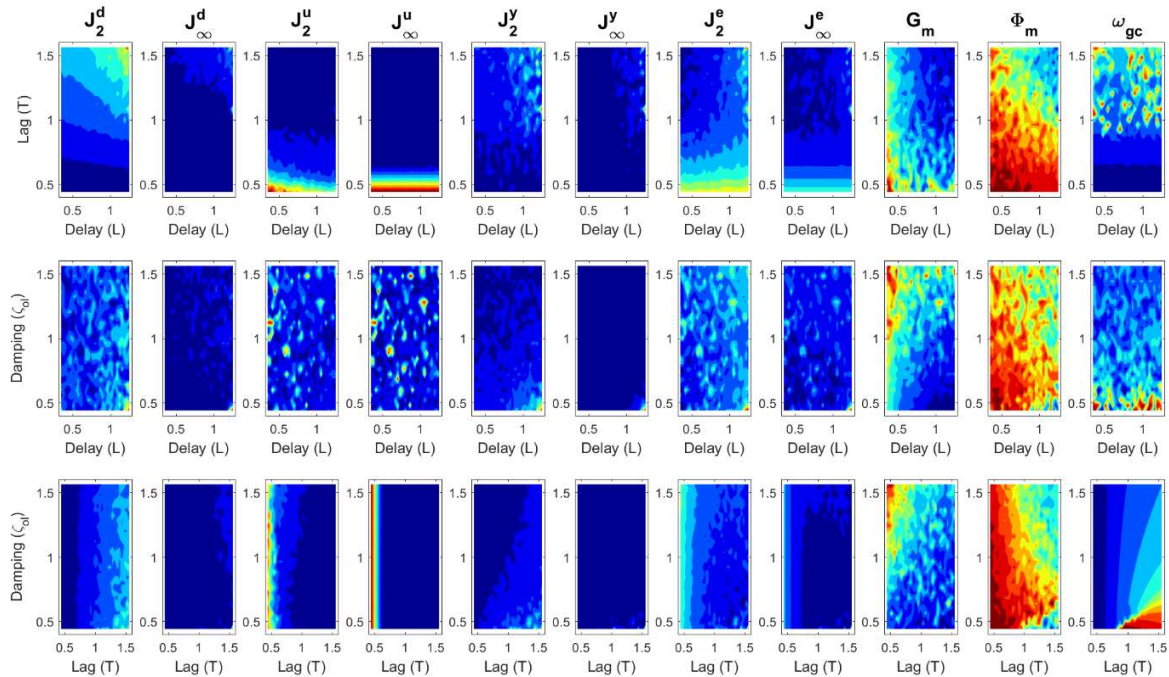

Figure 30: Iso-performance contours as joint distribution of 60% perturbed process parameters {L, T, $\zeta_{ol}$} with the robust stable PID controller having all real non-dominant poles for process $G_2$.

The 2D contour plots in Figure 29-Figure 37 for all the nine types of processes are explored with parameter variations around their respective robust stable controllers, designed with the nominal process parameters, using a perturbed range in the open loop parameters $\{L,T,\zeta_{ol}\}$. These are helpful to get an idea about how each process parameter affects the 11 different performance criteria, mentioned above. This can be justified from almost parallel lines either along the *x*-axis or *y*-axis which is found in several 2D sub-plots of certain process parameter pairs. For lag-dominant processes it can be observed that the performance criteria $\{G_m, \Phi_m, \omega_{gc}\}$ are more influenced by the delay (*L*) compared to the rest two process parameters $\{T, \zeta_{ol}\}$ as shown in Figure 29-Figure 31. The gain/phase margins $\{G_m, \Phi_m\}$ can be maintained to higher value (for better robust stability and performance) with smaller $\{L,T\}$ for the lag-dominant processes. Whereas gain crossover frequency ($\omega_{gc}$) can be increased to achieve faster time response with a lower value of damping ($\zeta_{ol}$) for the lag-dominant processes. Also, for the lag-dominant processes a lower value of delay helps maintaining a smaller 2/∞-norms.

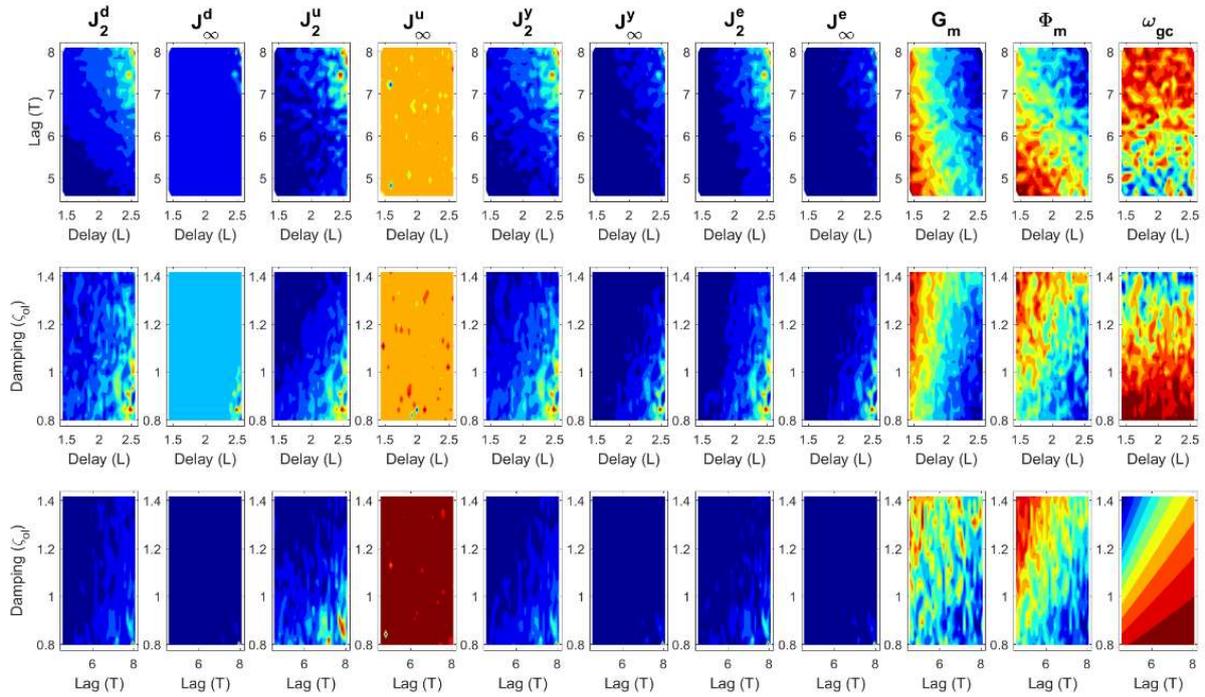

*Figure 31: Iso-performance contours as joint distribution of 60% perturbed process parameters {L, T, ζ$_{ol}$} with the robust stable PID controller having all real non-dominant poles for process G$_3$.*

For the balanced lag-delay processes ($G_4$-$G_6$) most of the performance criteria are highly influenced by the lag (*T*) compared to the rest two parameters $\{L, \zeta_{ol}\}$ which are observed from almost parallel lines along the axes with *T* in Figure 32-Figure 34. Also, a large phase margin ($\Phi_m$) can be maintained for low delay (*L*). Similar to the balanced lag-delay processes, the delay-dominant process ($G_7$-$G_9$) the lag parameter (*T*) has more influence to all the 11 performance criteria as shown in Figure 35-Figure 37, compared to the rest two process parameter $\{L, \zeta_{ol}\}$. Moreover, low delay (*L*) has got mild effect to achieve a higher phase margin ($\Phi_m$) and hence less overshoot for the delay-dominant processes.

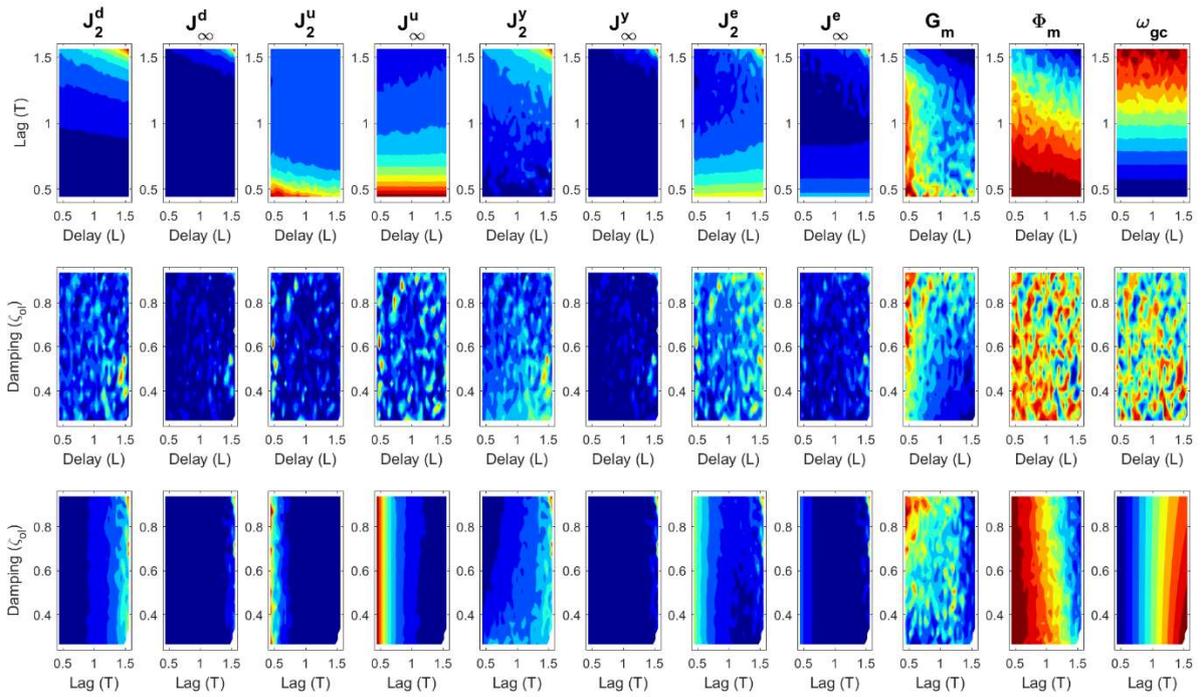

*Figure 32: Iso-performance contours as joint distribution of 60% perturbed process parameters {L, T, $\zeta_{ol}$} with the robust stable PID controller having all real non-dominant poles for process $G_4$.*

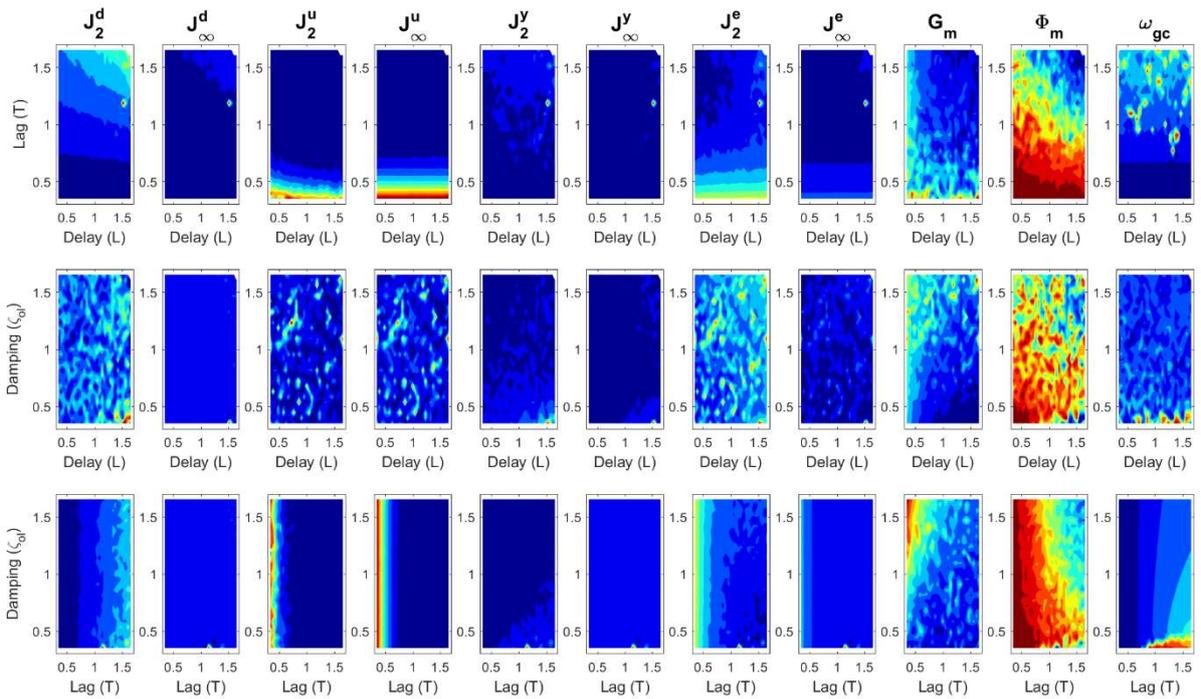

*Figure 33: Iso-performance contours as joint distribution of 70% perturbed process parameters {L, T, $\zeta_{ol}$} with the robust stable PID controller having all real non-dominant poles for process $G_5$.*

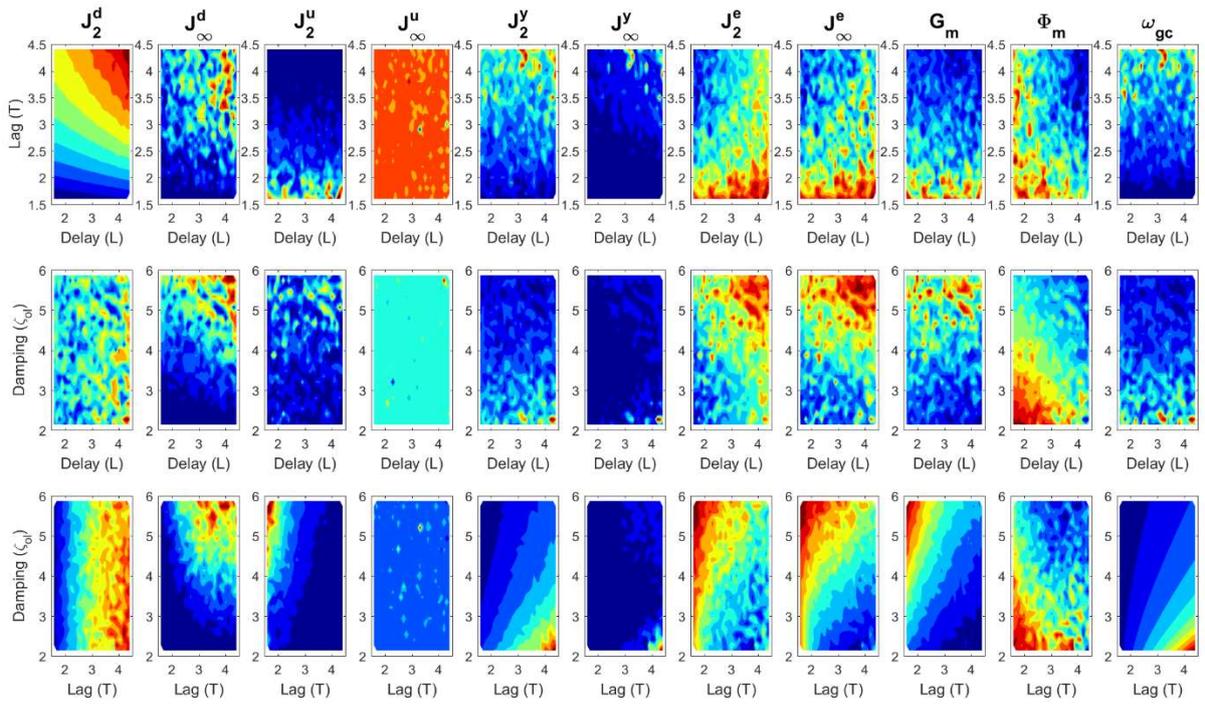

*Figure 34: Iso-performance contours as joint distribution of 50% perturbed process parameters {L, T, $\zeta_{ol}$} with the robust stable PID controller having all real non-dominant poles for process $G_6$.*

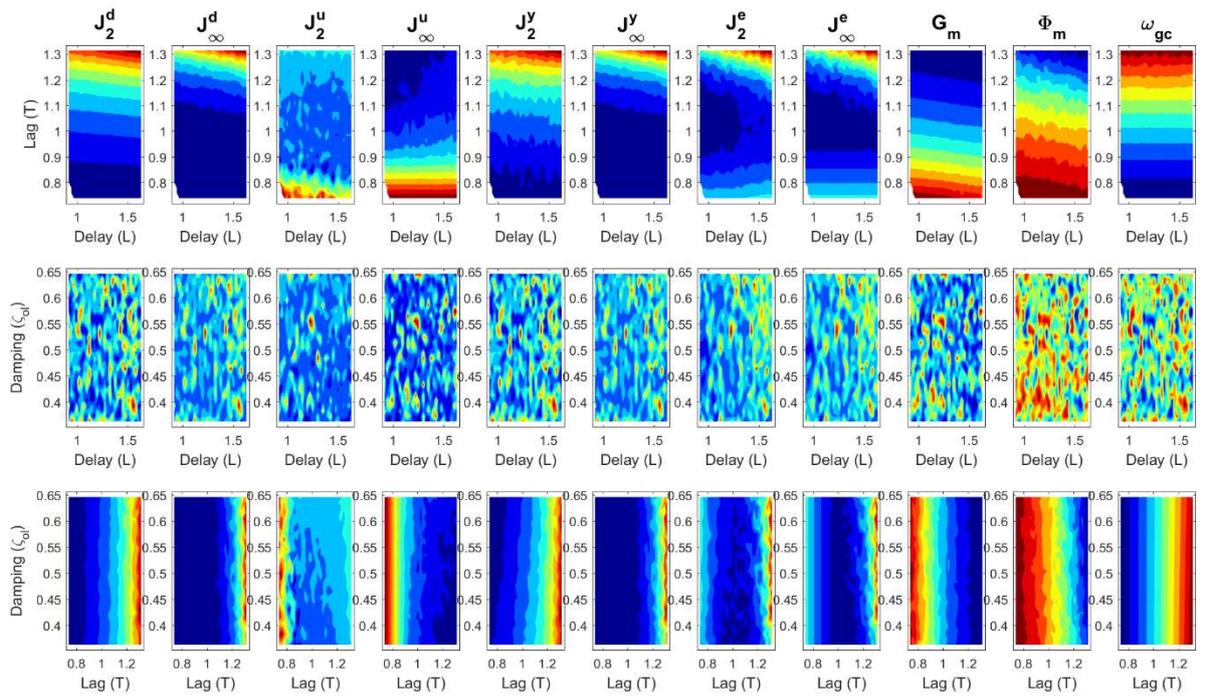

*Figure 35: Iso-performance contours as joint distribution of 30% perturbed process parameters {L, T, $\zeta_{ol}$} with the robust stable PID controller having all real non-dominant poles for process $G_7$.*

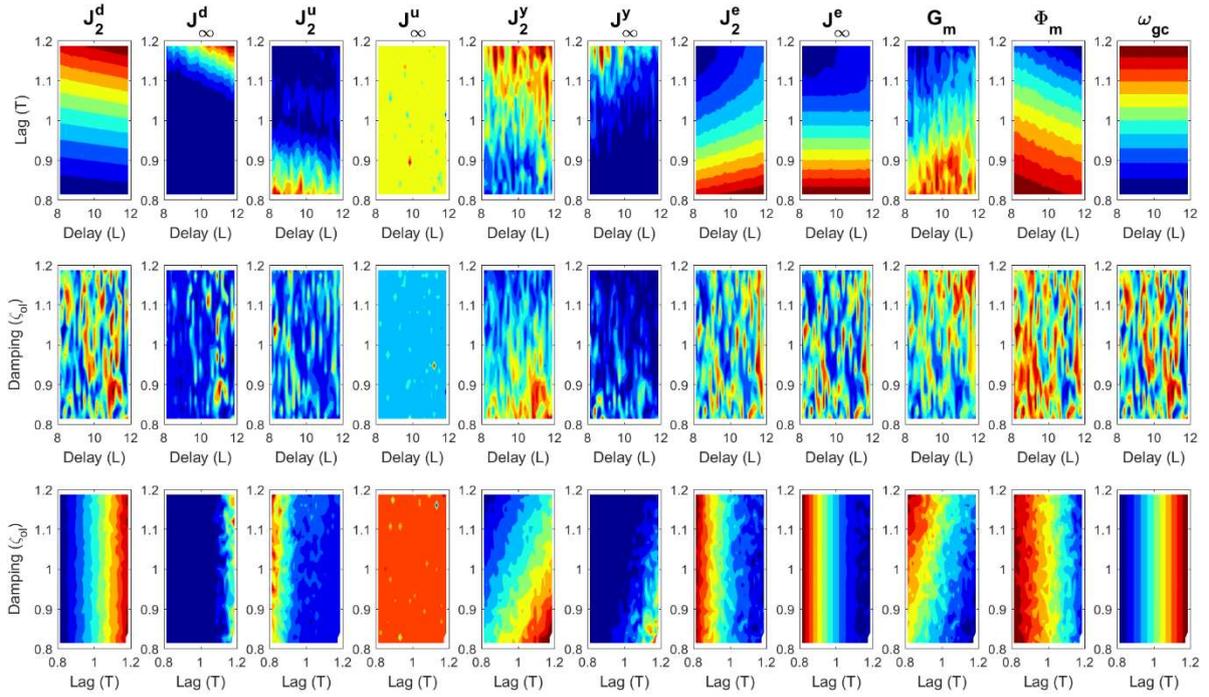

*Figure 36: Iso-performance contours as joint distribution of 20% perturbed process parameters {L, T, $\zeta_{ol}$} with the robust stable PID controller having all real non-dominant poles for process $G_8$.*

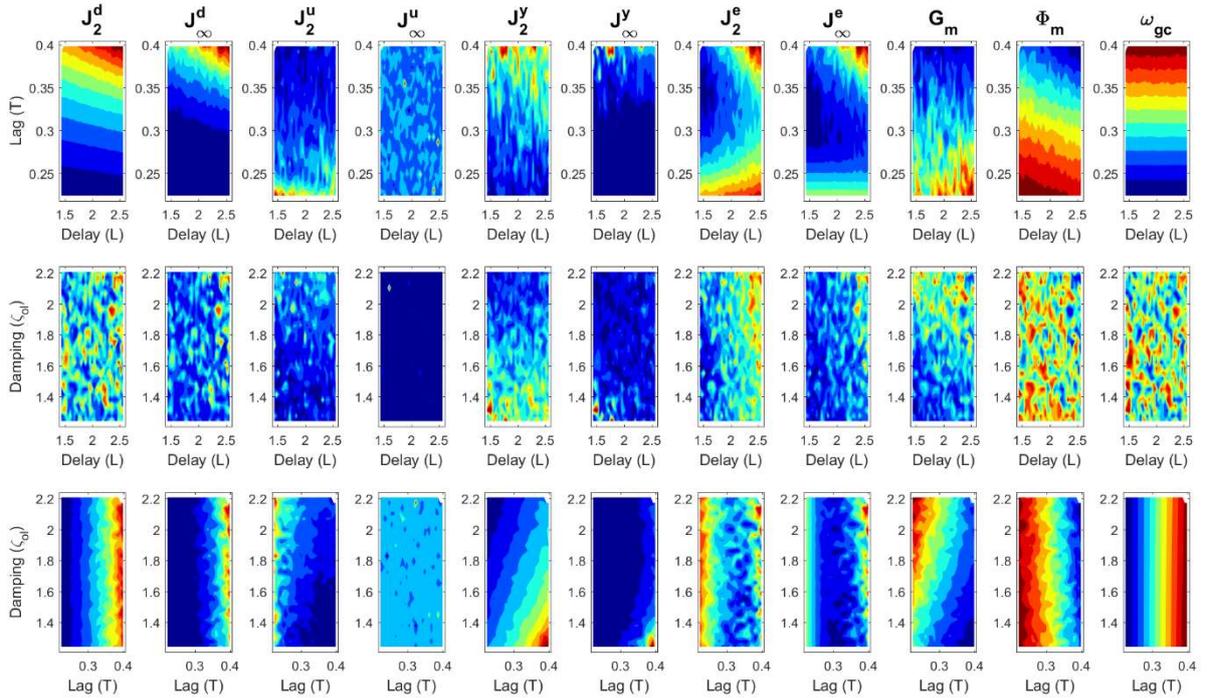

*Figure 37: Iso-performance contours as joint distribution of 30% perturbed process parameters {L, T, $\zeta_{ol}$} with the robust stable PID controller having all real non-dominant poles for process $G_9$.*

### 4.6. PID controller tuning rule generation for robust stability

The robust stable PID controller gains as the centroid of the stabilizing cluster of points for each test-bench processes can now be condensed in the form of polynomial functions or tuning rules involving the process parameters like delay to time constant ratio ($L/T$) and open loop damping ($\zeta_{ol}$), while following a similar method adopted in (Das et al. 2015) using polynomial regression analysis on the open loop process model parameters. However in this case the criteria is changed from several optimal performances to the robust stable solutions unlike (O'Dwyer 2009)(Das et al. 2015). The tuning rule generation for robust stability can be summarized as (36) for mapping the open loop process parameters in (25)-(33) on to the controller gains in Table 2. Owing to the lesser number of process models (9 plants) and in order to obtain a consistent estimate, the maximum order for $K_p$ tuning rule is limited to 2 in both the free parameters $\{L/T, \zeta_{ol}\}$ while for $\{K_i, K_d\}$ tuning rules, a second order polynomial lag to delay ratio ($L/T$) and first order polynomial in damping ($\zeta_{ol}$) are considered along with the interaction terms between them, as shown in (36).

$$\{K_p, K_i, K_d\} = f(L/T, \zeta_{ol})/K$$
$$K_p \sim p_{00} + p_{10}(L/T) + p_{01}\zeta_{ol} + p_{20}(L/T)^2 + p_{11}(L/T)\zeta_{ol} + p_{02}\zeta_{ol}^2 \quad (36)$$
$$\{K_i, K_d\} \sim p_{00} + p_{10}(L/T) + p_{01}\zeta_{ol} + p_{20}(L/T)^2 + p_{11}(L/T)\zeta_{ol}$$

Table 4: Model parameter for tuning rule generation and fitting statistics

| PID parameters | Order of $L/T$ | Order of $\zeta_{ol}$ | RMSE | $R^2$ | Adjusted $R^2$ |
|---|---|---|---|---|---|
| $K_p$ | 2 | 2 | 1.3263 | 0.8762 | 0.6698 |
| $K_i$ | 2 | 1 | 0.0706 | 0.9629 | 0.9258 |
| $K_d$ | 2 | 1 | 4.4964 | 0.6693 | 0.3386 |

This final model order selection (amongst many other combinations) to develop the optimum tuning rule for the PID controller gains satisfying all the nine test-bench plants was carried out using the adjusted coefficient of determination ($Adj\ R^2$) criteria as shown in Table 4. This is a preferred criterion since it penalizes both model complexity and fitting error, rather than using the simple criteria like root mean squared error (RMSE) and $R^2$ as both of them are prone to overfitting small datasets. It has been found that the best fitting performance ($Adj\ R^2 \approx 1$) is found for the integral gain $K_i$. The best fitted coefficients of the tuning rule (36) for the three PID controller gains are given in (37) along with their respective uncertainties.

$$\begin{aligned} K_p: \quad & p_{00} = 5.4815 \pm 2.648,\ p_{10} = -8.2986 \pm 2.2708,\ p_{01} = 1.7928 \pm 3.0842, \\ & p_{20} = 0.5475 \pm 0.14441,\ p_{11} = 2.1487 \pm 0.81942,\ p_{02} = -0.68766 \pm 0.59573; \\ K_i: \quad & p_{00} = 1.1271 \pm 0.10005,\ p_{10} = -0.64476 \pm 0.12037,\ p_{01} = -0.25487 \pm 0.045225, \\ & p_{20} = 0.036846 \pm 0.0076874,\ p_{11} = 0.19132 \pm 0.041707; \\ K_d: \quad & p_{00} = 17.247 \pm 6.3708,\ p_{10} = -17.696 \pm 7.6642,\ p_{01} = -2.8946 \pm 2.8796, \\ & p_{20} = 1.1552 \pm 0.48948,\ p_{11} = 4.8674 \pm 2.6557. \end{aligned} \quad (37)$$

### 4.7. Computational complexity of the algorithm

In the overall design method proposed in this paper, there are mainly two steps where the computational complexity needs to be considered:

i) Random Monte Carlo sampling to get the stability regions
ii) Using the $k$-means clustering algorithm on the stabilizing data points in the 3D space of PID controller gains

Theoretical computational complexity analysis for Monte Carlo algorithms may be quite involved (Troyer & Wiese 2005). We have therefore provided a numerical study of the computational complexity, similar to the work reported in (Karlsson et al. 2005) for Sequential Monte Carlo algorithms. A similar numerical simulation based approach has been adopted for the complexity analysis of the $k$-means clustering algorithm as shown in (Velmurugan & Santhanam 2010).

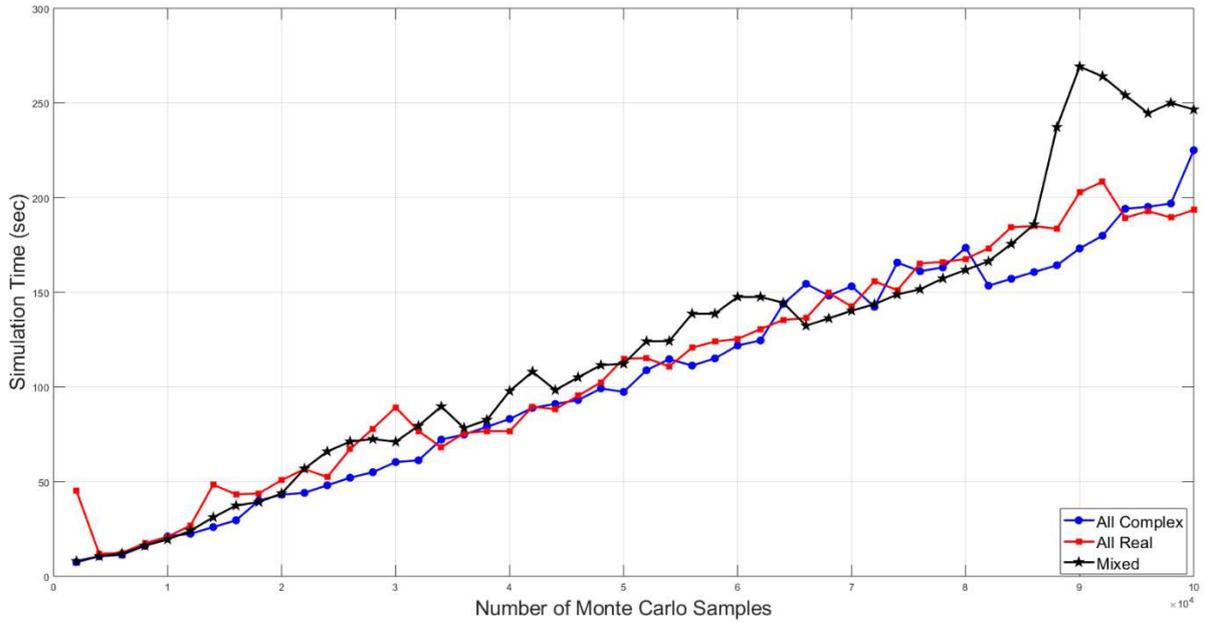

*Figure 38: Scalability of the Monte Carlo sampling with increasing sample number for plant $G_1$ and all three non-dominant pole types.*

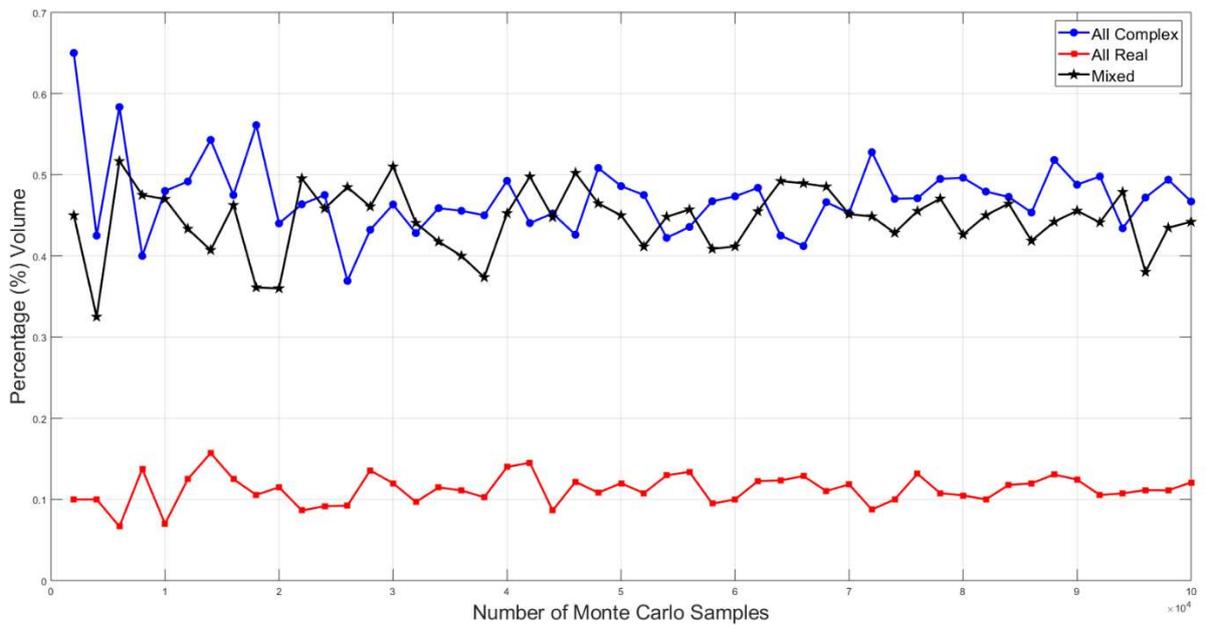

*Figure 39: Convergence of the percentage volume of the stability region with increasing sample number for plant $G_1$ and all three non-dominant pole types.*

The simulations were run on a 64-bit Windows PC with 16 GB memory and an Intel I5, 3.40 GHz processor on 4 parallel cores. The run times of the Monte Carlo sampling increases with the number of samples drawn within the chosen interval of the design parameters which are compared in Figure 38 for three different non-dominant pole types. The corresponding percentage volume of the stability region converges to their final value as the number samples increases which is shown in Figure 39. The *k*-means clustering algorithm also takes slightly longer time when finding the centroid in a larger volume of datasets which is also shown in Figure 40. It is important to note that here the sampling or clustering algorithms are employed in a 3D parameter space. Usually for the same dimension and number of data-points, increased number of compact clusters in the data-set may increase the computational time of the *k*-means algorithm (Velmurugan & Santhanam 2010). Similar to the previous cases, here as well 10 runs of the *k*-means algorithm were carried out and the best results with minimum distance criteria was selected in order to avoid the effect of randomized searching from different initial conditions.

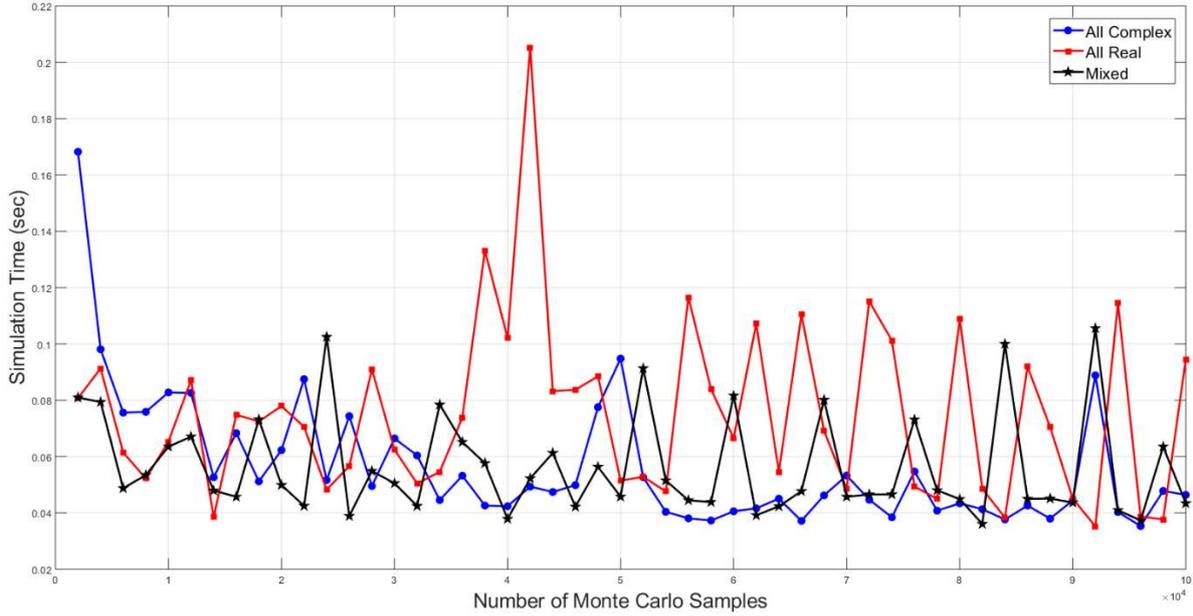

*Figure 40: Time requirements for the k-means clustering algorithm with increasing sample number for plant $G_1$ and all three non-dominant pole types.*

### 4.8. Hypothesis testing using the sampled data for different processes and non-dominant pole types

As shown in Table 1, there can be 27 possible combinations for different process types based on lag to delay ratios (lag-dominated, balanced and delay-dominated), damping levels (under-damped, critically-damped and over-damped) and non-dominant pole types (all complex, all real and mixed type). It would be interesting to study whether these groups have significant effect on the sampled data-points which represent the stability regions for each case. Here, the number of factors is 27 (different combinations of plant and non-dominant pole types) and there are 6 response variables ($K_p, K_i, K_d, \zeta_{cl}, \omega_{cl}, m$). The Multivariate Analysis of Variance (MANOVA) compares multivariate group means with two or more continuous response variables. This helps in avoiding the chance of inflated type-I error in group assignment compared to carrying out many univariate analysis of variance (ANOVA) tests on each response variable (Field et al. 2012). The simple ANOVA also would have ignored any relationship between the response variables which can be seen from the joint scatter plots reported in the earlier sections amongst the set of PID controller gains and the design parameters. Before, the MANOVA test was conducted, it is worthwhile to look at the underlying distributions of the datasets representing the stability regions in the joint parameter space of PID controller gains and three design variables. The group-wise scatter plots in Figure 41 show that there are some partial overlap between the 27 groups. However, the hypothesis testing will help us to answer the question whether the mean of these distributions are different or not.

For the MANOVA test, usually there are some assumptions about the nature of the dataset e.g. independence, multivariate normality and homogeneity of the covariance matrices. The normality of the datasets are violated as observed from the scatterplots in Figure 41 and also in the earlier sections as well as more formally from the Kolmogorv-Smirnov test on the univariate datasets out of the multivariate one ($K_p, K_i, K_d, \zeta_{cl}, \omega_{cl}, m$) or even a multivariate normality test. The independence condition is violated if the data comes from the same process which is the case here. Therefore, here we need to use the repeated measure MANOVA. Four different test statistics have been used to quantify repeated measure MANOVA results i.e. Pillai-Bartlett trace ($V$), Hotelling-Lawley trace ($U$), Wilk's lambda ($\Lambda$), Roy's largest root statistic ($\Theta$) as given in (38).

$$\begin{aligned}
V &= trace\left(Q_h \left(Q_h + Q_e\right)^{-1}\right) = \sum \theta_i, \\
\Lambda &= \frac{|Q_e|}{|Q_h + Q_e|} = \prod \frac{1}{1+\lambda_i}, \\
U &= trace\left(Q_h Q_e^{-1}\right) = \sum \lambda_i, \\
\Theta &= \max\left(eig\left(Q_h Q_e^{-1}\right)\right).
\end{aligned} \tag{38}$$

In (38), the hypothesis sum of squares and products matrix ($Q_h$) and the residuals sum of squares and products matrix ($Q_e$) are given as:

$$\begin{aligned}
Q_h &= T^T Z^{-1} T \\
Q_e &= C^T \left(R^T R\right) C,
\end{aligned} \tag{39}$$

where, $R = Y - X\hat{B}$.

The other parameters in (38) are defined as:

$$\begin{aligned}
T &= A\hat{B}C - D \\
Z &= A\left(X^T X\right)^{-1} A^T,
\end{aligned} \tag{40}$$

with $ABC = D$, $B$ is the matrix of coefficients, $A$ is the matrix defining hypothesis based on the between subject models, $C$ is the matrix defining hypotheses based on the within-subjects model and $D$ is the matrix containing hypothesized value. More details on the multivariate hypothesis testing can be found in (Davis 2002).

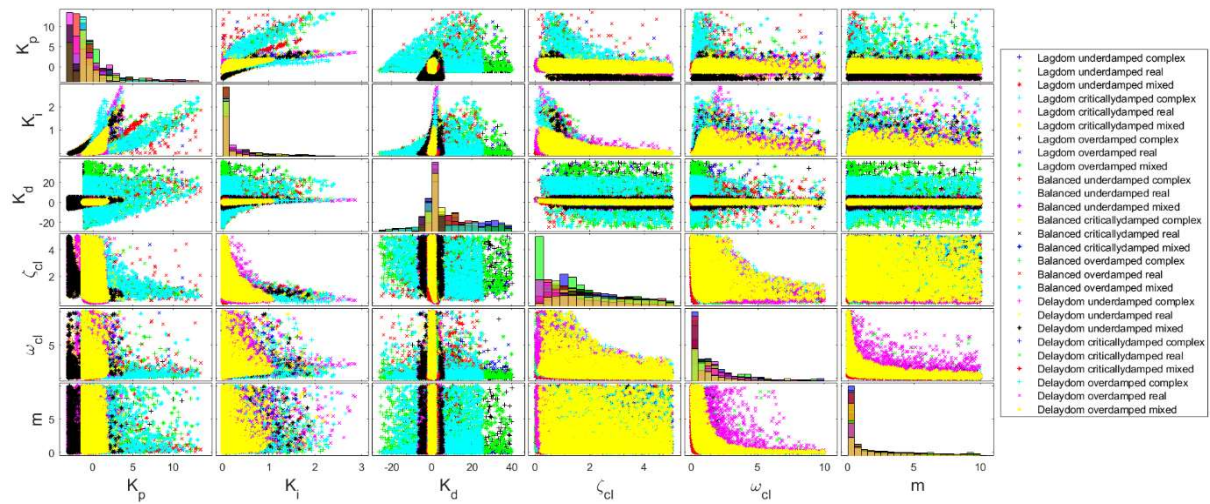

Figure 41: Group-wise bivariate scatter plots for different process and non-dominant pole types.

It has been suggested in (Field et al. 2012), that even though the conditions like multivariate normality or homogeneity of the covariance matrices within the groups are violated, some measures can still be useful like the Pillai-Bartlett trace. However the Roy's largest root statistic may be affected by the non-normality (especially platykurtic distributions) as well as non-homogeneity of the covariance matrices as discussed in (Field et al. 2012). The repeated measure MANOVA results are given in Table 5 which shows that for all the different statistics, the group differences are indeed significant ($p<0.01$).

Table 5: Repeated measure MANOVA results for the 27 groups representing different process dynamics (lag to delay ratio and damping) and non-dominant pole types

| Within | Between | Statistic | Value | F Statistic | $R^2$ | df1 | df2 | p-value |
|---|---|---|---|---|---|---|---|---|
| Constant | Intercept | Pillai | 0.76717 | 20238 | 0.76717 | 5 | 30710 | 0 |
| Constant | Intercept | Wilks | 0.23283 | 20238 | 0.76717 | 5 | 30710 | 0 |
| Constant | Intercept | Hotelling | 3.2951 | 20238 | 0.76717 | 5 | 30710 | 0 |
| Constant | Intercept | Roy | 3.2951 | 20238 | 0.76717 | 5 | 30710 | 0 |
| Constant | group | Pillai | 0.82663 | 233.98 | 0.16533 | 130 | 153570 | 0 |
| Constant | group | Wilks | 0.35719 | 270.48 | 0.18856 | 130 | 151310 | 0 |
| Constant | group | Hotelling | 1.3332 | 314.92 | 0.2105 | 130 | 153540 | 0 |
| Constant | group | Roy | 0.93328 | 1102.5 | 0.48274 | 26 | 30714 | 0 |

However, with an assumption of independence, as such each data-point came from different plants and there being no relation between the data-points which were actually sampled from the same plant, we can use the standard one-way MANOVA. The one-way MANOVA results for all the 6 observed variables show significant results ($p<0.01$) for comparing 27 groups with the respective $p$-values being $p = \left[0, 0, 0, 0, 4.7 \times 10^{-251}, 8.05 \times 10^{-38}\right]$ with degrees of freedom $d = 6$.

Table 6: Results of univariate Kruskal-Wallis tests on each variable

| Variable | $\chi^2$ statistic | p-value |
|---|---|---|
| $K_p$ | 10132.8 | 0 |
| $K_i$ | 3784.994 | 0 |
| $K_d$ | 3963 | 0 |
| $\zeta_{cl}$ | 3265.832 | 0 |
| $\omega_{cl}$ | 8316.116 | 0 |
| $m$ | 2388.299 | 0 |

It is common to do a post-hoc analysis after a significant MANOVA test, by carrying out several univariate ANOVA tests with Bonferroni correction by diving the significance level with the number of tests conducted. However due the violation of the normality condition, here we carry out the non-parametric version of ANOVA i.e. the Kruskal-Wallis test. This test uses the null hypothesis that each of the 6 parameters $\{K_p, K_i, K_d, \zeta_{cl}, \omega_{cl}, m\}$ between different groups come from the same distribution whereas the alternative hypothesis being that not all the samples come from the same distribution. The respective $\chi^2$-statistics (instead of the $F$-statistics in standard one way ANOVA) and $p$-values have been reported in Table 6 for each of the six controller gain and design parameters, along with the group-wise box-plots in Figure 42 which clearly shows that their medians are different as verified from the hypothesis tests. With the Bonferroni correction, these six univariate non-parametric tests have shown significant results ($p<0.01$) for all the variables amongst $\{K_p, K_i, K_d, \zeta_{cl}, \omega_{cl}, m\}$, indicating significant group differences amongst the 27 possible combinations. The significantly low *p*-values show that the tests reject the null hypothesis indicating a significant group difference.

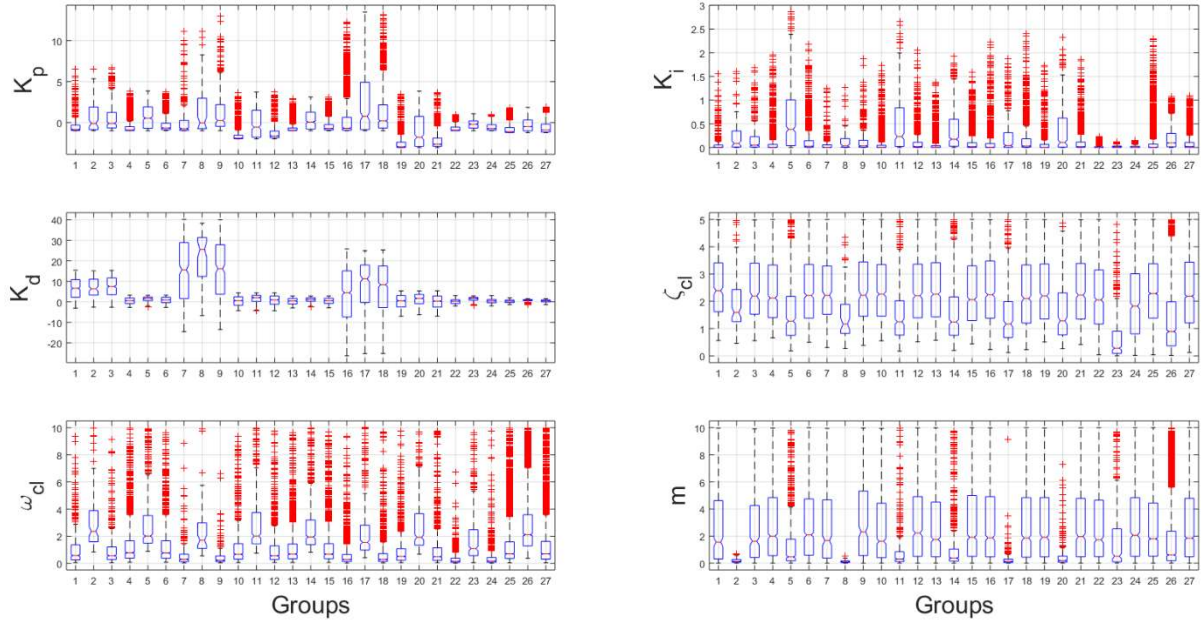

*Figure 42: Univariate box-plots for {$K_p$, $K_i$, $K_d$, $\zeta_{cl}$, $\omega_{cl}$, m} as given by the Kruskal-Wallis hypothesis test.*

5. Discussions

This paper proposes three types of dominant pole placement based PID controller design methods for processes with time delays using different criteria for the non-dominant poles i.e. all complex conjugate, all real and mixture of them. The assumption of different types of non-dominant poles are used to get many alternative candidate expressions for the PID controller gains, so that the designer can choose the most robust stable one amongst competing solutions, while also maintaining acceptable closed loop performance, characterised by 11 different control performance measures which is followed by a correlation analysis amongst them. Although the present design was primarily focussed on dominant pole placement technique, depending on the nature of the dominant poles the closed loop performance may vary drastically. Also, the dominant pole placement parameters $\{m, \zeta_{cl}, \omega_{cl}\}$ include the designer's choice of the expected level of non-dominance, closed loop oscillation and speed of operation respectively. For a range of demanded closed loop parameters $\{m, \zeta_{cl}, \omega_{cl}\}$, there can be many alternative stabilizing controller gains which can form a stability region in the controller parameter space which was found out by randomly sampling the specified interval of design parameters. The robust stable solution is chosen as the centroid of the stability region for different constructs of the characteristic equations considering either all real, all complex or their combinations for the closed loop non-dominant poles. However, as a side effect of such robust stable design, while satisfying the dominant pole placement criteria, the other control performance objectives like servo/regulatory, tracking/disturbance rejection, control effort, noise rejection trade-offs in the form of various system/signal norms as well as traditional frequency domain measures like gain/phase margin, gain cross-over frequency might take some arbitrary values, rendering unacceptable or poor overall closed loop performance. It is well known that all these performance measures cannot be directly controlled in either the pole placement or robust stable design approaches but may be of prime concern in many industrial process control applications. Therefore our design approach first finds out the centroid of the stabilizing PID controller gains, thus maximizing parametric robustness of the process model, since uncertainty in the linear system parameters can be equivalently represented as uncertainty in the PID controller parameters (Silva et al. 2007). Then with the best robust stable solutions which can be judged either by comparing the highest volume of the stability region or some other time/frequency domain performance criteria, we investigate rest of the performance measure as an additional exploration. However, incorporating multiple of these desired criteria can be better managed under a unified or conflicting control design within a linear or bilinear matrix inequality (LMI/BMI) approach or by using a multi-objective optimisation which can be investigated in a future research.

The overdetermined nature of the PID controller design problem having 4 alternative solutions for $K_p$ in different non-dominant pole-types produce different volume and shape of the stability regions. Amongst them the largest stability region as indicated by the maximum number of accepted stable solutions within the same interval of design parameters, is selected. Then *k*-means clustering algorithm is used to find out the robust stable PID controller gains within the largest stability region. It was observed from the previous sections that the robust stable PID controller which has been designed considering the all real non-dominant pole criteria performs the best compared to the other non-dominant pole types i.e. all complex conjugate and the mixture of them. Again, the robust stable PID controller, designed with all real non-dominant poles, is also capable to handle the SOPTD processes with large parametric uncertainties around their nominal values where this uncertainty has been considered in the process parameters $\{L, T, \zeta_{ol}\}$, instead of the controller gains. Also, simple tuning rules has been established *via* polynomial regression analysis to design the PID controller gains to achieve robust stability using the nine types of test-bench SOPTD processes. Computational complexity analysis of the two step algorithm has been provided, followed by hypothesis testing for the distribution of the sampled data-points for different process and non-dominant pole types.

However, in spite of demonstrating the effectiveness of the proposed PID controller design methodology to handle the stable test-bench SOPTD processes e.g. performance invariance against order of time delay approximation using Pade method, there are also some limitations of this approach:

- The design has not considered the effect of any strong zero or lead dynamics in the time delay processes. As the presence of the time delay term in the SOPTD process itself makes the system to have an infinite order (upon infinite series expansion of the exponential term), the inclusion of the zero dynamics in the delayed process makes it even more difficult to design dominant pole placement criteria based PID controller which can be considered as our future scope of research.
- Performance contours and stability regions are obtained by random sampling but not through any deterministic analytical expression.
- This design methodology assumes about prior knowledge of the plant model, although in a more realistic case the plant models can be uncertain either being norm bounded or having varying parameters within certain intervals. For such a scenario, the robust stable PID controller design usually should not be based on the dominant pole placement criteria on the nominal system and is left as future scope of research.

Regarding the main achievement of this paper, we first attempt to not only find the stability region of an SOPTD process which can be thought of a higher order system under Pade approximation of the delay term but also, we can maintain a desired equivalent delay-free second order response, characterised by a user-specified closed loop damping and frequency. In other words, the effect of undesired poles and zeros due to the Pade approximation of delay term can be constrained with the non-dominance criteria (controlled by the parameter *m*) such that even with arbitrary order of $N_{Pade}$ and hence unknown number of poles and zeros in the closed loop system, the dominant process dynamics can still be maintained near to the desired level which has been successfully shown in Figure 25-Figure 27. It is a new finding on top of the existing dominant pole placement PID controller design related literatures mentioned before, since our method is insensitive of the actual number of non-dominant poles/zeros, arising from arbitrary order of Pade approximation for the delay term and thus both the servo/regulatory modes can be maintained at an acceptable level for such time-delay systems *via* a simpler dominant pole placement approach.

## 6. Conclusion

In order to handle SOPTD processes, dominant pole placement based PID controller tuning has been considered with three different non-dominant pole types. The *k*-means clustering technique is used here to achieve the robust stable PID controller gains inside the stability region in the controller parameter space, obtained by random Monte Carlo sampling of the chosen pole placement parameter intervals. Different closed loop performance measures in both time and frequency domain have been analysed for nine test-bench SOPTD processes with their corresponding robust tuned PID controller which was originally designed using analytical criteria derived for a third order Pade approximation ($N_{Pade}$ = 3). It has been shown later that the proposed robust-stable PID controllers are able to control the resulting higher order processes and dynamics of arbitrary number of poles and zeros, due to expansion of the delay term i.e. increasing the order of the Pade approximation while maintaining a similar closed loop performance. In future, the proposed methodology can be extended for nonlinear and uncertain processes. Also, to design a digital control scheme, the proposed methodology can be further extended in the discrete time version where choice of the sampling time as a fraction of desired closed loop time constant may play a significant role.

**Acknowledgement**

KH acknowledges the support from the University Grants Commission (UGC), Govt. of India under its Basic Scientific Research (BSR) scheme.